\documentclass[prd,preprint,superscriptaddress,amsmath,amssymb,nofootinbib]{revtex4}
\pdfoutput=1

\usepackage{graphicx,color,slashed}% Include figure files

\def\be{\begin{eqnarray}}
\def\ed{\end{eqnarray}}

\begin{document}

%{\begin{flushright}{xxx}
%\end{flushright}}

\title{\bf \Large Light charged Higgs boson with dominant decay to
quarks and its search at LHC and future colliders}

\author{A.G. Akeroyd}
\email{a.g.akeroyd@soton.ac.uk}
\affiliation{School of Physics and Astronomy, University of Southampton,
Highfield, Southampton SO17 1BJ, United Kingdom}

\author{Stefano Moretti}
\email{S.Moretti@soton.ac.uk}

\affiliation{School of Physics and Astronomy, University of Southampton,
Highfield, Southampton SO17 1BJ, United Kingdom}

\author{Muyuan Song}
\email{ms32g13@soton.ac.uk}

\affiliation{School of Physics and Astronomy, University of Southampton,
Highfield, Southampton SO17 1BJ, United Kingdom}

\date{\today}% It is always \today, today,

\begin{abstract}
The possibility of a light charged Higgs boson $H^\pm$ that decays 
predominantly to quarks ($cs$ and/or $cb$) and with a mass in the
range 80 GeV $\le m_{H^\pm} \le 90$ GeV
is studied in the context of Three-Higgs-Doublet Models (3HDMs).
At present the Large Hadron Collider (LHC) has little sensitivity to this 
scenario, and currently the best constraints are from LEP2 and Tevatron searches. 
The branching ratio of $H^\pm\to cb$ can be dominant in two of the five types of 3HDM, and
we determine the parameter space where this occurs. The decay  $H^\pm\to cb$ has recently been searched for at the 
LHC for the first time, and with increased integrated luminosity one would expect sensitivity
to the region 80 GeV $\le m_{H^\pm} \le 90$ GeV due to the smaller backgrounds with respect
to $H^\pm\to cs$ decays.

\end{abstract}

\maketitle

\section{Introduction}
\noindent

In 2012 the ATLAS and CMS collaborations of the Large Hadron Collider (LHC) 
announced the discovery of a new particle with a mass of around 125 GeV
\cite{Aad:2012tfa,Chatrchyan:2012xdj}. 
The current measurements of its properties are in very good agreement 
(within experimental error) with those of the Higgs boson of the 
Standard Model (SM), and measurements suggest that it has a spin of zero. 
Five decay channels ($\gamma\gamma$, $ZZ$, $WW$, $\tau\tau$, and $bb$) 
have now been observed with a statistical significance of greater than 
$5\sigma$ (e.g. see \cite{Aaboud:2018zhk}). 
The measured branching ratios (BRs) are in agreement with 
those predicted for the SM Higgs boson. Moreover, the main four 
production mechanisms (gluon-gluon fusion, vector boson $(W/Z)$ 
fusion, associated production with a vector boson, and 
associated production with top quarks) have been observed, 
with no significant deviation from the cross-sections of the SM Higgs boson.

The simplest assumption is that the observed 125 GeV boson is the 
(solitary) Higgs boson of the SM. However, it is possible that 
it is the first scalar to be discovered from a non-minimal Higgs 
sector, which contains additional scalar isospin doublets 
or higher representations such as scalar isospin triplets. 
In such a scenario, future measurements of the BRs of the 125 GeV boson
could show deviations from 
those of the SM Higgs boson. There is also the possibility of 
discovering additional neutral scalars, or physical charged scalars ($H^\pm$). 
In the context of a Two-Higgs-Doublet Model (2HDM) 
the lack of observation of 
an $H^\pm$ at the LHC rules out parameter space of $\tan\beta$ 
(from the Yukawa coupling) and $m_{H^\pm}$, where $\tan\beta=v_2/v_1$, 
and $v_1$ and $v_2$ are the vacuum expectation values (VEVs) of the two 
Higgs doublets respectively (for reviews see e.g. 
\cite{Branco:2011iw,Akeroyd:2016ymd}). 

In a Three-Higgs-Doublet Model (3HDM) the Yukawa couplings of the two 
charged scalars depend on the four free parameters ($\tan\beta$, $\tan\gamma$, 
$\theta$, and $\delta$) of the unitary matrix
that rotates the charged scalar fields in the weak eigenbasis to the
physical charged scalar fields.  As pointed out in previous works 
\cite{Grossman,Akeroyd:1994ga,Akeroyd:1995cf,Akeroyd2,Akeroyd:2016ssd}, 
in a 3HDM there is a phenomenologically attractive 
possibility of an $H^\pm$ being light ($m_{H^\pm} < m_t$) 
{\it and} having a large BR for the decay channel $H^\pm \to cb$,
a scenario which would not be expected in a 2HDM with 
natural flavour conservation (NFC) \cite{Glashow:1976nt} 
due to the stringent bounds
from the decay $b\to s \gamma$. 
A search for $H^\pm\to cb$ decays originating from $t\to H^\pm b$
has recently been performed at the LHC 
\cite{Sirunyan:2018dvm}.
The only study of the BRs of the two $H^\pm$s in 3HDMs (with NFC) as 
functions of the above four parameters was in Ref.~\cite{Akeroyd:2016ssd}. 
However, this work did not fully study the 
dependence of the BRs on the parameter space. We perform the first 
comprehensive study of the BRs of the lightest $H^\pm$ in the various 3HDMs 
(with NFC) as a function of the four parameters. We also study the dependence of
the product BR($t\to H^\pm b)\times$BR($H^\pm\to cb$), which gives the
number of events in the search in \cite{Sirunyan:2018dvm}.
We give emphasis to the scenario of 80 GeV$\le  m_{H^{\pm}} \le 90$ GeV and a 
large BR$(H^\pm \to cs/cb)$ for which detection is currently challenging at the LHC, but 
prospects with the anticipated integrated luminosities are more promising.

This work is organised as follows. 
In section II we give an introduction to the phenomenology 
of the lightest $H^\pm$ in 3HDMs with NFC. 
In section III the searches for $H^\pm$ 
at past and present colliders that 
provide sensitivity to the region 
$80\;{\rm GeV}< m_{H^\pm}<90$ GeV are summarised.
In section IV our results are presented, and conclusions are contained in
section V.

\section{The Three-Higgs-Doublet Model (3HDM) with NFC}
In this section the fermionic couplings of the lightest $H^\pm$ in the 3HDM 
as a function of the parameters of the scalar potential 
are presented. The constraints on the fermionic couplings are summarised, 
and explicit formulae for the BRs of 
the decay of $H^\pm$ to fermions are given.

\subsection{Fermionic couplings of $H^\pm$ in the 3HDM}
In a 2HDM the Lagrangian that corresponds to the interactions of $H^\pm$ with 
the fermions (the Yukawa couplings) can be written as follows:
\begin{equation}
{\cal L}_{H^\pm} =
-\left\{\frac{\sqrt2V_{ud}}{v}\overline{u}
\left(m_d X{P}_R+m_u Y{P}_L\right)d\,H^+
+\frac{\sqrt2m_\ell }{v} Z\overline{\nu_L^{}}\ell_R^{}H^+
+{H.c.}\right\}\,.
\label{lagrangian}
\end{equation}
Here $u(d)$ refers to the up(down)-type quarks, and $\ell$
refers to the electron, muon and tau.
The imposition of NFC, which eliminates tree-level flavour changing neutral
currents (FCNCs) that are mediated by scalars, leads to four distinct 2HDMs \cite{Barger}: Type I, Type II, lepton-specific, and
flipped. In Table~\ref{couplings} the couplings $X$, $Y$, and $Z$ in the 
four distinct
2HDMs are given.
\begin{table}[h]
\begin{center}
\begin{tabular}{|c||c|c|c|}
\hline
& $X$ &  $Y$ &  $Z$ \\ \hline
Type I
&  $-\cot\beta$ & $\cot\beta$ & $-\cot\beta$ \\
Type II
& $\tan\beta$ & $\cot\beta$ & $\tan\beta$ \\
Lepton-specific
& $-\cot\beta$ & $\cot\beta$ & $\tan\beta$ \\
Flipped
& $\tan\beta$ & $\cot\beta$ & $-\cot\beta$ \\
\hline
\end{tabular}
\end{center}
\caption{The couplings $X$, $Y$, and $Z$ in the Yukawa interactions of $H^\pm$ in the four versions of the 2HDM with NFC.}
\label{couplings}
\end{table}
The Lagrangian in eq.~(\ref{lagrangian}) also applies to the lightest $H^\pm$ 
of a 3HDM, with
the $X$, $Y$, and $Z$ couplings being functions of four parameters 
of a unitary matrix $U$.
This matrix $U$ connects the charged scalar fields in the 
weak eigenbasis ($\phi^\pm_1,\phi^\pm_2,\phi^\pm_3)$ with the 
physical scalar fields
($H^\pm_1$, $H^\pm_2$) and the charged Goldstone boson $G^\pm$ as follows:

\begin{equation}
	\left( \begin{array}{c} G^+ \\ H_2^+ \\ H_3^+ \end{array} \right) 
	= U \left( \begin{array}{c} \phi_1^+ \\ \phi_2^+ \\ \phi_3^+ \end{array} \right).
	\label{eq:Udef}
\end{equation}
We take $H^\pm_1$ as the lighter
of the two charged Higgs bosons, and from now on it is referred to as $H^\pm$
with the following couplings \cite{Logan}: 
\begin{equation}
	X = \frac{U_{d2}^\dagger}{U_{d1}^\dagger}, \quad \quad 
	Y = - \frac{U_{u2}^\dagger}{U_{u1}^\dagger}, \quad \quad 
	Z = \frac{U_{\ell 2}^\dagger}{U_{\ell 1}^\dagger}\,.
\label{eq:xyz}
\end{equation}
The values of $d$, $u$, and $\ell$ in these matrix elements  are given in Table \ref{valuesudl} and 
depend on which of the five distinct 3HDMs is under consideration. 
Taking $d=1$, $u=2$, and $\ell=3$ means that the down-type quarks receive their mass from the vacuum expectation value $v_1$,
the up-type quarks from $v_2$, and the charged leptons from $v_3$ (this choice is called the ``democratic 3HDM''). The other
possible choices of $d$, $u$, and $\ell$ in a 3HDM are given the same names as the four types of 2HDM.
The couplings of the $H^\pm_2$ (i.e. the heavier charged scalar) are obtained from eq.~(\ref{eq:xyz}) by
making the replacement $2 \to 3$ in the numerators of $X$, $Y$, and $Z$. We will not study these couplings 
for $H^\pm_2$ because our
focus will be on $H^\pm$ in the range 80 GeV$< m_{H^\pm} < 90$ GeV.

The matrix $U$ can be written explicitly as a function of
four parameters  $\tan\beta$, $\tan\gamma$, $\theta$, and $\delta$, where
\begin{equation}
	\tan\beta = v_2/v_1, \qquad \tan\gamma = \sqrt{v_1^2 + v_2^2}/v_{3}\,.
\end{equation}
and $v_1$, $v_2$, and $v_3$ are the VEVs.
The angle $\theta$ and phase $\delta$ can be written explicitly as  
functions of several parameters in the scalar potential \cite{Logan}.
The explicit form of $U$ is:
\begin{eqnarray}
	U &=& \left( \begin{array}{ccc} 
		1 & 0 & 0 \\
		0 & e^{-i \delta} & 0 \\
		0 & 0 & 1 \end{array} \right)
		\left( \begin{array}{ccc}
		1 & 0 & 0 \\
		0 & c_\theta & s_\theta e^{i \delta} \\
		0 & -s_\theta e^{-i \delta} & c_\theta \end{array} \right)
		\left( \begin{array}{ccc}
		s_\gamma & 0 & c_\gamma \\
		0 & 1 & 0 \\
		-c_\gamma & 0 & s_\gamma \end{array} \right)
		\left( \begin{array}{ccc}
		c_\beta & s_\beta & 0 \\
		-s_\beta & c_\beta & 0 \\
		0 & 0 & 1 \end{array} \right)
	\nonumber \\
	&=& \left( \begin{array}{ccc}
	s_\gamma c_\beta & s_\gamma s_\beta 	& c_\gamma \\
	-c_\theta s_\beta e^{-i\delta} - s_\theta c_\gamma c_\beta 
		& c_\theta c_\beta e^{-i\delta} - s_\theta c_\gamma s_\beta & s_\theta s_\gamma \\
	s_\theta s_\beta e^{-i\delta} - c_\theta c_\gamma c_\beta 
		& -s_\theta c_\beta e^{-i\delta} - c_\theta c_\gamma s_\beta & c_\theta s_\gamma 
	\end{array} \right).
	\label{eq:Uexplicit}
\end{eqnarray}
Here $s$ and $c$ denote the sine or cosine of the respective angle. Hence the functional forms of
$X$, $Y$, and $Z$ in a 3HDM depend on four parameters. This is in contrast to the analogous couplings in 
the 2HDM for which $\tan\beta$ is the only free coupling parameter.

\begin{table}[h]
\begin{center}
\begin{tabular}{|c||c|c|c|}
\hline
& $u$ &  $d$ &  $\ell$ \\ \hline
3HDM (Type I) &  2 & 2 & 2 \\
3HDM (Type II) & 2 & 1 & 1 \\
3HDM (Lepton-specific) & 2 & 2 & 1 \\
3HDM (Flipped) & 2 & 1 & 2 \\
3HDM (Democratic) & 2 & 1 & 3 \\
\hline
\end{tabular}
\end{center}
\caption{The five versions of the 3HDM with NFC,
and the corresponding values of $u$, $d$, and $\ell$. Taking $u=2$ means that the up-type quarks receive their mass
from the vacuum expectation value $v_2$, and likewise for $d$ (down-type quarks) and $\ell$ (charged leptons).}
\label{valuesudl}
\end{table}

\subsection{Constraints on the couplings $X$, $Y$, and $Z$}
The couplings $X$, $Y$, and $Z$ (and their combinations) are constrained
from various low-energy processes. Detailed studies in the context
of the Aligned 2HDM (for which the couplings of $H^\pm$ are
also given by $X$, $Y$, and $Z$) can be found in Refs.~\cite{Jung:2010ik,Trott:2010iz}. These constraints can be
applied to the lightest $H^\pm$ of a 3HDM provided that the 
contribution to a given process from the $H^\pm_2$ is considerably 
smaller (e.g. if $m_{H^\pm_2}\gg m_{H^\pm}$). In this work we assume that 
any contribution from $H^\pm_2$ is sub-dominant and can be neglected to a 
good approximation.
We summarise here the bounds (which are also summarised in \cite{Logan}) 
that we will use in our numerical analysis.

The coupling $Y$ is constrained from the process $Z\to b\overline b$ 
from LEP data. For $m_{H^\pm}$ around 100 GeV (on which we focus) 
the constraint is roughly $|Y|<1$ (assuming $|X|\le 50$, so that the
dominant contribution is from the $Y$ coupling).  
The coupling $X$ is also constrained from $Z\to b\overline b$, but the 
constraints from this process are weaker than those from $t\to H^\pm b$
(which will be studied later in this work).

From the rare decay $b\to s\gamma$ a constraint 
on the combination Re$(XY^*)$ is given by
\begin{equation}
-1.1 \le {\rm Re}(XY^*)\le 0.7.
\label{bsgamma}
\end{equation}
This constraint was derived in \cite{Trott:2010iz} for  
$m_{H^\pm}=100$ GeV, and is an approximation for the case when 
i) the contribution from $|Y|^2$ can be neglected (which
is a fairly good approximation because $|Y|<1$) and ii)
Im($XY^*$) is small (which is a good approximation, as shown shortly below).
Detailed constraints on the $H^\pm$ contribution
to $b\to s \gamma$ in the Aligned 2HDM without this
approximation can be found in \cite{Jung:2010ik}. 
Other works are usually in the 
context of the 2HDM with NFC 
\cite{Ciuchini1,Ciuchini2,Borzumati,Gambino,Misiak,Misiak2}. 

In a 3HDM one would have contributions to 
$b\to s\gamma$ from both $H^\pm$ and $H^\pm_2$. 
The only study of the prediction for BR($b\to s\gamma$) in 3HDMs 
to next-to-leading order accuracy is in \cite{Akeroyd:2016ssd}.
It was shown there that there exists
parameter space for which $H^\pm$ can be of the order of 80 GeV
even for Type II and flipped structures (which would not be possible
in the 2HDM with these structures). This is due to the 
additional presence of $H^\pm_2$
and the larger number of parameters in the couplings $X$ and $Y$
with respect to the 2HDM with NFC. 
In our numerical analysis for the BRs of $H^\pm$ 
we will use the allowed range given in
eq.~(\ref{bsgamma}) in order to find the regions of $\tan\beta$, $\tan\gamma$, 
$\theta$, and $\delta$ that satisfy the $b\to s\gamma$ constraint. Although
eq.~(\ref{bsgamma}) neglects the contribution from $H^\pm_2$ we will 
take eq.~(\ref{bsgamma}) as being representative of the $b\to s\gamma$ 
constraint in 3HDMs. The true region allowed by $b\to s\gamma$
(to next-to leading order accuracy, as done in  \cite{Akeroyd:2016ssd}) 
would presumably be shifted somewhat from 
the regions allowed by eq.~(\ref{bsgamma}). We argue later that we
would not expect this to significantly alter our qualitative results.

The electric dipole moment of the neutron gives the following constraint
on Im$(XY^*)$ \cite{Trott:2010iz}:
\begin{equation} 
|{\rm Im}(XY^*)|\le 0.1\,.
\label{Imxy}
\end{equation}
This bound is for $m_{H^\pm}=100$ GeV and is an order-of-magnitude
estimate. There are also the constraints $|Z|\le 40$ and $|XZ|\le 1080$, both
for $m_{H^\pm}=100$ GeV. In our numerical analysis we will respect all these constraints.

\subsection{The Branching Ratios of $H^\pm$}
We will only consider the decays of $H^\pm$ to fermions. If there exists a neutral scalar
(e.g. a CP-even $h^0$ or a CP-odd $A^0$) that is lighter than $H^\pm$ then the
decay channel $H^\pm \to h^0W^*$ and/or $H^\pm\to A^0W^*$ would be open and can be sizeable
(or even dominant) \cite{Akeroyd2,Moretti:1994ds,Djouadi:1995gv,Akeroyd:1998dt,Kling:2015uba,Arhrib:2016wpw,Arbey:2017gmh,Arhrib:2017wmo}. 
We assume that these decays are negligible, and
this is most easily achieved by taking $m_{A^0}, m_{h^0}>m_{H^\pm}$.
 In a 3HDM the expressions for the partial widths of the decay modes to fermions
of $H^\pm$ are:

\begin{equation}
\Gamma(H^\pm\to \ell^\pm\nu)=\frac{G_F m_{H^\pm} m^2_\ell |Z|^2}{4\pi\sqrt 2}\;, 
\label{width_tau}
\end{equation}
\begin{equation}
\Gamma(H^\pm\to ud)=\frac{3G_F V_{ud}m_{H^\pm}(m_d^2|X|^2+m_u^2|Y|^2)}{4\pi\sqrt 2}\;.
\label{width_ud}
\end{equation}

In the expression for 
$\Gamma(H^\pm\to ud)$ the running quark masses should be evaluated at the 
scale of $m_{H^\pm}$, and
there are QCD vertex corrections which multiply the partial widths by
$(1+17\alpha_s/(3\pi))$. 
A study of the BRs as a function of $|X|$, $|Y|$, and $|Z|$ 
was first given in \cite{Akeroyd:1994ga} and more recently in \cite{Akeroyd2}.
For $|X|\gg |Y|,|Z|$ the decay channel BR$(H^\pm\to cb)$ can dominate
(which was first mentioned in \cite{Grossman}), reaching
a maximum of $\sim 80\%$. In contrast, in a 2HDM with NFC the only model
which contains a 
parameter space for a large BR$(H^\pm\to cb)$ with $m_{H^\pm}< m_t$
is the flipped model (a possibility mentioned in \cite{Grossman,Akeroyd:1994ga}
and studied in more detail in \cite{Aoki:2009ha,Logan:2010ag}), 
However, in this case the $b\to s\gamma$ constraint would require
$m_{H^\pm}>500$ GeV \cite{Misiak2} for which $H^\pm\to tb$ would dominate.

The first study of the dependence of the BRs of $H^\pm$ in 3HDMs in terms
of the parameters $\tan\beta, \tan\gamma, \theta,$ and $\delta$ was given in 
\cite{Akeroyd:2016ssd}. 
However, this work did not fully study the dependence of the BRs 
on the parameter space (i.e. $\delta=0$, $\theta=-\pi/4$, and
$\tan\beta=2(5)$ was taken as a representative choice), and showed the BRs as a function of $\tan\gamma$ only. Moreover,
in \cite{Akeroyd:2016ssd} the dependence of the BRs on the model parameters was carried out in the Higgs basis, and
so the parameters $\tan\beta, \tan\gamma, \theta$, and $\delta$ used in that work are not equivalent to
the corresponding parameters in this work.  

We now briefly mention other models in which a large BR($H^\pm\to cb$) is possible, although in this work we
will just study the 3HDMs with NFC.
The $X,Y,$ and $Z$ couplings of
$H^\pm$ in the Aligned 2HDM \cite{Pich:2009sp} (which does not have NFC, but instead eliminates scalar FCNCs at tree level by taking 
certain Yukawa matrices to be proportional to each other) are functions of five parameters. 
Consequently,  $|X|\gg |Y|,|Z|$ can be realised and a large BR($H^\pm\to cb$) is possible \cite{Akeroyd2}. In the 2HDM 
(Type III) in which fermions receive their masses from both VEVs (and
scalar FCNCs are present at tree level), the Yukawa couplings of $H^\pm$ depend on more parameters 
than in the Aligned 2HDM and thus a large  BR($H^\pm\to cb$) can be obtained \cite{HernandezSanchez:2012eg}.
Similar comments apply to a Four-Higgs-doublet model \cite{Logan}. In models for which $X, Y,$ and $Z$ 
depend on several parameters one expects some parameter
space for a large BR$(H^\pm\to cb)$ for $m_{H^\pm} < m_t$, while satisfying the $b\to s \gamma$ constraint.  
\\

\section{Searches for $H^\pm$ in the 
region 80 GeV $\le m_{H^\pm}\le 90$ GeV}
We focus on the scenario of $H^\pm$ being lighter than the top quark.  
There have been searches for $H^\pm$ in the region 80 GeV $\le m_{H^\pm}\le 
90$ GeV at LEP2, Tevatron and the LHC. However, the sensitivity to this mass region is often inferior to
that for 90 GeV$\le m_{H^\pm}\le 160$ GeV because of the large backgrounds from $W$ decays.
We pay particular attention to the region of 80 GeV $\le m_{H^\pm}\le 
90$ GeV, and in the following we discuss the searches for 
$m_{H^\pm}<m_t$ at each of these colliders.

\subsection{Tevatron Searches}
At the Fermilab Tevatron the production mechanism is 
$p\overline p\to t\overline t$, where one top quark decays
conventionally via $t\to Wb$ and 
the other top quark decays via $t\to H^\pm b$.
Taking $|V_{tb}|=1$ one has the following expressions for 
the decays of a top quark to a $W$ boson or an $H^\pm$:
\begin{eqnarray}
\Gamma(t\to W^\pm b)=\frac{G_F m_t}{8\sqrt 2 \pi}[m_t^2+2M_W^2][1-M_W^2/m_t^2]^2\,,  \\ \nonumber
\Gamma(t\to H^\pm b)=\frac{G_F m_t}{8\sqrt 2 \pi}[m^2_t|Y|^2 + m_b^2|X|^2][1-m_{H^\pm}^2/m_t^2]^2\,.
\end{eqnarray}
As can be seen from the above equations the BR($t\to H^\pm b$) depends on 
the magnitude of $|X|$ and $|Y|$. As discussed earlier, 
the BRs of $H^\pm$
depend on the {\it relative} values of $|X|,|Y|$ and $|Z|$. The search by the D0 collaboration
in \cite{Abazov:2009aa} with 1 fb$^{-1}$ of data obtained the following 
limit in the region  80 GeV $\le m_{H^\pm}\le 
90$ GeV:
\begin{equation}
{\rm BR}(t\to H^\pm b) < 0.21\;\;{\rm for}\;\;50\% \le{\rm BR}(H^\pm\to cs)\le 100\%\,.
\end{equation}
In the search strategy in \cite{Abazov:2009aa} 
the presence of a large BR($H^\pm\to cs/cb$) 
in the decay $t\to H^\pm b$ would lead to a 
depletion in the expected number of events in the $\ell$ +jets, $\ell\ell$ and
$\ell\tau$ channels ($\ell=e$ or $\mu$) compared to that expected from
$t\overline t\to W^+W^-b\overline b$. Importantly, this ``disappearance'' 
search has sensitivity to the region 80 GeV $\le m_{H^\pm}\le 
90$ GeV and is thus an effective strategy 
when ${\rm BR}(H^\pm\to cs/cb)$ is large and $m_{H^\pm}$ lies in
the above region.

The CDF collaboration (with 2.2 fb$^{-1}$) used a different search strategy \cite{Aaltonen:2009ke}
in which the signature of $H^\pm \to cs$
was searched for as a peak at $m_{H^\pm}$ in the invariant mass distribution of the quarks that
it decays to (i.e. an ``appearance'' search for $H^\pm\to cs$). This
technique provides limits on BR$(t\to H^\pm b)$ 
that are competitive with those in \cite{Abazov:2009aa} for 
values of $m_{H^\pm}$ that are not
in the region 80 GeV $\le m_{H^\pm}\le 90$ GeV. However the search provides no constraints
for $80\;{\rm GeV}\le m_{H^\pm}\le 90$ GeV because the background
from $W\to q\overline q$ is too large.
Up to now the LHC has only carried out appearance searches for 
$H^\pm \to cs/cb$ (see below).

\subsection{LHC Searches}

The production mechanism at the LHC is $pp\to t\overline t$, 
where one top quark decays via $t\to H^\pm b$ (i.e. the same mechanism at the parton level
as at the Tevatron). The LHC is expected to have accumulated around $150$ fb$^{-1}$
of integrated luminosity at $\sqrt s=13$ TeV by the end of the year 2018, at which point long shut down
2 will commence. Various searches for the decay
$t\to H^\pm b$ have been carried out at the LHC, and are summarised in Table~\ref{LHC_search}.

\begin{table}[h]
\begin{center}
\begin{tabular}{|c||c|c|}
\hline
& ATLAS &  CMS  \\ \hline
7 TeV (5 fb$^{-1}$)
&  $cs$ \cite{Aad:2013hla}, $\tau\nu$ \cite{Aad:2012rjx,Aad:2012tj}
&  $\tau\nu$ \cite{Chatrchyan:2012vca} \\
8 TeV (20 fb$^{-1}$)
& $\tau\nu$ \cite{Aad:2014kga} & $cs$ \cite{Khachatryan:2015uua}, 
$cb$ \cite{Sirunyan:2018dvm}, 
$\tau\nu$ \cite{Khachatryan:2015qxa}  \\
13 TeV (36 fb$^{-1}$)
& $\tau\nu$ \cite{Aaboud:2018gjj} & $\tau\nu$ \cite{CMS:2016szv,CMS:2018ect}  \\
\hline
\end{tabular}
\end{center}
\caption{Searches for $H^\pm$ at the LHC, using $pp\to t\overline t$ and $t\to H^\pm b$. The given integrated
luminosities are approximate. The search in \cite{Chatrchyan:2012vca}
used 2 fb$^{-1}$, and the search in \cite{CMS:2016szv} used 
13 fb$^{-1}$.}
\label{LHC_search}
\end{table}

\subsubsection{Decay $H^\pm \to \tau\nu$}
For the decay $H^\pm\to \tau\nu$ there are four basic signatures, which arise from  
the leptonic and hadronic decays of $H^\pm$ and $W^\pm$. Searches for three of these signatures
have been carried out with the 7 TeV data \cite{Chatrchyan:2012vca,Aad:2012rjx,Aad:2012tj}, which were then combined to give
a limit on the product BR$(t\to H^\pm b)\times {\rm BR}(H^\pm\to \tau\nu$) for a given $m_{H^\pm}$. Note that ATLAS used
two different search strategies  \cite{Aad:2012rjx,Aad:2012tj} that give comparable sensitivity.
In \cite{Chatrchyan:2012vca,Aad:2012rjx,Aad:2012tj} the  limit is roughly $\ge 4\%$ for $m_{H^\pm}=90$ GeV, 
which strengthens with increasing $m_{H^\pm}$ to $\ge 1\%$ for $m_{H^\pm}=160$ GeV. Only the CMS search \cite{Chatrchyan:2012vca}
presented limits ($\ge 4\%$) for the mass range 80 GeV $\le m_{H^\pm}\le 90$ GeV.

In the searches for $H^\pm\to \tau\nu$ 
with the 8 TeV data \cite{Khachatryan:2015qxa,Aad:2014kga} both the $\tau$ and the $W$ boson from $t\to W^\pm b$ decay were taken
to decay hadronically. This signature (of the four) offers the greatest sensitivity
at present. The transverse mass of $H^\pm$
is calculated from its decay products of hadrons and missing energy.
Both the ATLAS and CMS searches presented limits for the  mass range 80 GeV $\le m_{H^\pm}\le 90$ GeV.
Limits on the product BR$(t\to H^\pm b)\times {\rm BR}(H^\pm\to \tau\nu$)
were obtained, being around $\ge 1\%$ for $m_{H^\pm}=80$ GeV
and strengthening with increasing $m_{H^\pm}$ to $\ge 0.2\%$ for $m_{H^\pm}=160$ GeV.

The CMS search  \cite{CMS:2016szv} with 13 TeV data and 13 fb$^{-1}$
also used the hadronic decay of the $\tau$ from $H^\pm\to \tau\nu$, and
selected the hadronic decay of the $W^\pm$. Similar limits  
to those in \cite{Aad:2014kga} and \cite{Khachatryan:2015qxa}
were obtained, but are
slightly weaker for the region 80 GeV $\le m_{H^\pm} \le 90$ GeV.
Recently a CMS search was carried out with 13 TeV data and 36 fb$^{-1}$ 
\cite{CMS:2018ect}, 
which combined separate searches for three of the four basic signatures
(the case where both the $W$ and $\tau$ decay leptonically was not searched
for). Significantly improved limits on  
BR$(t\to H^\pm b)\times {\rm BR}(H^\pm\to \tau\nu$) 
were obtained, ranging from $\ge 0.36\%$ for $m_{H^\pm}=80$ GeV to
$\ge 0.08\%$ for $m_{H^\pm}=160$ GeV.

There has been a search with the 13 TeV data \cite{Aaboud:2018gjj}
from the ATLAS collaboration using 36 fb$^{-1}$, with
limits similar to those in  \cite{CMS:2018ect}. In contrast to the ATLAS search
with 8 TeV data \cite{Aad:2014kga}, both the leptonic and hadronic decays of the $W^\pm$ boson were considered
(the $\tau$ is still taken to decay hadronically). No limits are presented 
for the region 80 GeV $\le m_{H^\pm} \le 90$ GeV, but the sensitivity to $m_{H^\pm}>90$ GeV
has improved by a factor of approximately 5 to 10 e.g. for $m_{H^\pm}=90$ GeV the limit on 
 BR$(t\to H^\pm b)\times {\rm BR}(H^\pm\to \tau\nu$) is $\ge 0.3\%$, and with the 8 TeV data it is $\ge 1.2\%$.

\subsubsection{Decay $H^\pm \to cs/cb$}
ATLAS carried out a search for $H^\pm\to cs$ \cite{Aad:2013hla} with 5 fb$^{-1}$ of data at 7 TeV, 
while CMS \cite{Khachatryan:2015uua} carried out a search for $H^\pm\to cs$ 
using 20 fb$^{-1}$ of data at 8 TeV. The $W$ boson
is taken to decay leptonically. Two tagged $b-$quarks are required (which arise from the decay of the
$t-$quarks), and the invariant mass distribution of the two quarks that are not $b-$tagged
(i.e. the $c$ and $s$ quarks that
originate from $H^\pm$) is plotted. The signature of $H^\pm$ would be a peak at $m_{H^\pm}$ in this invariant mass
distribution. Limits
on the product BR$(t\to H^\pm b)\times {\rm BR}(H^\pm\to cs$) are obtained, 
which range from around $\ge 5\%$ for $m_{H^\pm}= 90$ GeV to  $2\%$ for  $m_{H^\pm}=160$ GeV.
Note that these limits are weaker than those for $H^\pm\to \tau\nu $ decay
for a given $m_{H^\pm}$.
In the invariant mass distribution the dominant background from 
$W\to qq$ decays gives rise to a peak around 80 GeV. 
Hence the expected sensitivity starts to weaken significantly with decreasing $m_{H^\pm}$ 
in the region 90 GeV $\le m_{H^\pm}\le 100$ GeV, and there are no limits for the region 80 GeV $\le m_{H^\pm} \le 90$ GeV.

CMS carried out a search  \cite{Sirunyan:2018dvm} for $H^\pm \to cb$ decays (assuming a branching ratio of $100\%$)
with the leptonic decay of $W$.
Signal events will have three $b-$quarks, although one (or more) might not be tagged as a $b-$quark.
Two event categories were defined: i) $3b+e^\pm$, and ii) $3b+\mu^\pm$.
%Due to the high $b$ tagging efficiencies $(70\%)$, most signal events will fall in the $3b$ categories.
A fitting procedure was carried out in order to correctly identify the tagged 
$b-$quark that arises from $H^\pm\to cb$,
which is then used (together with the non-$b$-tagged $c$ quark) in the invariant mass distribution of $H^\pm$. Due to BR($W\to cb$) being very small, the background to $H^\pm\to cb$ decays is much smaller than that
for $H^\pm\to cs$. Combining both event categories results in limits on BR$(t\to H^\pm b)\times {\rm BR}(H^\pm\to cb$)
of around $\ge 1.4\%$ for $m_{H^\pm}=90$ GeV, which strengthens 
with increasing $m_{H^\pm}$ to $\ge 0.5\%$ for $m_{H^\pm}=150$ GeV. 
These limits are stronger than those for $H^\pm \to cs$ decays for a given $m_{H^\pm}$.
Again, no limits are given in the mass range 80 GeV $\le m_{H^\pm}\le 90$ GeV, although 
(unlike the case for $H^\pm\to cs$) the sensitivity
does not diminish considerably in the range 90 GeV $\le m_{H^\pm}\le 100$ GeV.

\subsubsection{Sensitivity to 80 GeV $\le m_{H^\pm}\le 90$ GeV for future LHC searches for $H^\pm\to cs/cb$ }
Given the significantly lower backgrounds for the $3b$ signature 
arising from  $H^\pm\to cb$ decays it is hoped that future searches (e.g. with
150$^{-1}$ fb and $\sqrt s=13$ TeV) will be able to set limits 
on BR$(t\to H^\pm b)\times {\rm BR}(H^\pm\to cb$)
in the region 80 GeV $\le m_{H^\pm}\le 90$ GeV. Eventually, one would also expect some sensitivity in this region for the search with the
$2b$ signature (which is sensitive to $H^\pm\to cs/cb$ decays)
with 150$^{-1}$ fb and above. However, the limits would (most likely) be inferior to those in the $3b$ channel for a given luminosity.

As mentioned earlier, the Tevatron strategy of a disappearance search for
$H^\pm\to cs/cb$ has not yet been attempted at the LHC. A dedicated disappearance search at the LHC would be likely to 
improve on the Tevatron limit on BR$(t\to H^\pm b)\times {\rm BR}(H^\pm\to cs/cb$) of
$20\%$ \cite{Abazov:2009aa} for 80 GeV $\le m_{H^\pm}\le 90$ GeV. 
However, we are not aware of any LHC simulations, and so at present it is not clear whether or not this strategy
could give a sensitivity that is competitive with that for the appearance searches.

\subsection{LEP2 Searches and future $e^+e^-$ colliders}
The production mechanism at LEP2 was $e^+e^-\to H^+H^-$. An important
difference with the searches for $H^\pm$ at hadron colliders is that
the couplings $X,Y,Z$ do not appear in the production cross-section
for $e^+e^-\to H^+H^-$, which is instead a function of just one 
unknown parameter $m_{H^\pm}$. Hence this production mechanism 
at $e^+ e^-$ colliders can produce $H^\pm$ even with very 
small values of $X,Y,Z$, provided that $2m_{H^\pm} < \sqrt s$.

The LEP working group combined the separate searches from
the four LEP experiments \cite{Abbiendi:2013hk}. These searches
were carried out at energies in the range $\sqrt s=183$
GeV to $\sqrt s=209$ GeV, and with a total combined integrated luminosity
of 2.6 fb$^{-1}$. 
In the searches for the fermionic decay modes of $H^\pm$ it is assumed that BR$(H^\pm \to cs)$+BR$(H^\pm \to \tau\nu)=1$,
but the actual experimental search for $H^\pm\to cs$ would be also be sensitive to $H^\pm\to cb$ and other light flavours
of quark. Dedicated searches for the decay mode $H^\pm\to A^0W^*$ were also carried out in \cite{Abbiendi:2013hk}, but in this work we are assuming that
this channel is absent or very suppressed.
From the search for fermionic decays the excluded region at 95\% confidence level (CL)  
in the plane $[m_{H^\pm}, {\rm BR}(H^\pm\to \tau\nu)]$ is shown.
For $m_{H^\pm}<80$ GeV the whole range 
$0 \le {\rm BR}(H^\pm\to \tau\nu)\le 100\%$ is excluded. For $80\; {\rm GeV}\le \;m_{H^\pm}<90$ GeV, most of the region is not excluded
for BR$(H^\pm \to \tau\nu)<80\%$ (i.e.  BR$(H^\pm \to cs)>20\%$). Notably, there
is an excess of events of greater than $2\sigma$ significance around the point $m_{H^\pm}=89$ GeV, BR$(H^\pm \to cs)=65\%$ and BR$(H^\pm \to \tau\nu)=35\%$,
which could be readily accommodated in a 3HDM with appropriate choices
of $X$, $Y$ and $Z$.
As mentioned in our earlier work \cite{Akeroyd:2016ssd}
an excess like this is an example of a possible signal for $H^\pm$ that was 
just out of the range of LEP2. Such an excess, if genuine, 
could be observed at the LHC  provided that the values of $|X|$ and 
$|Y|$ are large enough to ensure enough events of $t\to H^\pm b$ at a given
integrated luminosity. 
Future LHC searches in the $\tau\nu$ channel, which
currently have sensitivity to the region 80 GeV $\le m_{H^\pm} \le 90$ GeV,
could then observe such an $H^\pm$. One could also expect a signal 
in the $H^\pm\to cs/cb$ channel provided that sensitivity to 
the region 80 GeV $\le m_{H^\pm} \le 90$ GeV is obtained. If
$|X|$ and $|Y|$ are sufficiently small then such an 
$H^\pm$ would escape detection
at the LHC, but could be observed at future $e^+e^-$ colliders (see below).

The possibility of a 
future circular $e^+e^-$ collider operating at a variety of energies from $\sqrt s=m_Z$ to $\sqrt s=2m_t$ is being discussed
(FCC-ee at CERN and CEPC in China), and a future $e^+e^-$ Linear Collider (ILC) would also take data in this energy range 
(and higher energies). If such machines are approved, the earliest starting date of operation for CEPC (FCC-ee)
would be the year 2030 (2040), with the ILC possibly starting between these two dates.
The choice of $\sqrt s=240$ GeV would be optimal for detailed studies of the discovered 125 GeV neutral boson. This
energy would also enable pair production of $H^\pm$ up to a mass of 120 GeV.
The integrated luminosity with $\sqrt s=240$ GeV at all three colliders is 
expected to be of the order of a few ab$^{-1}$, which is three orders of magnitude greater
than the integrated luminosity (2.6 fb$^{-1}$) used in the combined LEP search in \cite{Abbiendi:2013hk}.
Hence an $H^\pm$ with a mass in the region $80\; {\rm GeV}\le \;m_{H^\pm}<90$ GeV would be discovered
for any value of BR$(H^\pm \to cs/cb)$, with a signal in at least one 
of the three channels $H^+H^-\to jjjj, jj\tau\nu,\tau\nu\tau\nu$ (where $j$ signifies quarks lighter than the
$t$ quark). As mentioned earlier,  
the production mechanism $e^+e^-\to H^+H^-$ does not depend on the couplings to fermions. Hence
an $H^\pm$ with $2m_{H^\pm} < \sqrt s$ that escaped detection at the LHC due to small values
of $X,Y,$ and $Z$ would be discovered at the above $e^+e^-$ colliders.

\section{Results}
We vary the four input parameters that determine $X$, $Y$, and $Z$ 
in the following ranges
(see e.g. \cite{Logan}):
\begin{eqnarray}
-\pi/2\le \theta \le 0,\;\;\;\;\;\;\;\;\;\;  0\le \delta \le 2\pi,   
\nonumber \\
1 \le \tan\beta \le 60, \;\;\;\;\; 1 \le \tan\gamma \le 60 \,.
\label{4param}
\end{eqnarray}
We have checked that the phenomenological constraints on $|X|$, $|Y|$, $|Z|$,
and $|XZ|$ from section II.B are respected but we do not show explicit plots
for these parameters. The constraints on Re$(XY^*)$ and 
Im$(XY^*)$ rule out significant regions of parameter space, and these
will be shown in specific plots. Taking 
$\delta=0$ leads to real values for $X,Y,$ and $Z$, and so 
in this case the constraint on 
Im($XY^*$) will be automatically respected. 
We only consider $m_{H^\pm}< m_t$,
and results will be presented for the cases of $m_{H^\pm}=85$ GeV (for which the LHC currently has no sensitivity if BR$(H^\pm\to cs/cb)$ is dominant) and
$m_{H^\pm}=130$ GeV (for which the LHC has already set limits). We pick $m_{H^\pm}=85$ GeV as a representative choice that is
midway in the interval 80 GeV$\le m_{H^\pm}\le 90$ GeV, although our results will
apply to all values of $m_{H^\pm}$ in this interval with small numerical differences.
The searches at LEP \cite{Abbiendi:2013hk} cannot rule out BR$(H^\pm\to cs)\ge 50\%$
for 80 GeV$\le m_{H^\pm}\le 83$ GeV and 88 GeV$\le m_{H^\pm}\le 90$ GeV.
However, in the interval 83 GeV$\le m_{H^\pm}\le 88$ GeV 
the values BR$(H^\pm\to cs)\ge 90\%$ are ruled out (but $50\% \le {\rm BR}(H^\pm\to cs)\le 90\%$ are not). 
We do not impose this small excluded region on our figures.

In our numerical analysis we are only concerned with the 
four parameters in eq.~(\ref{4param}) and $m_{H^\pm}$. 
These comprise five of the sixteen\footnote{There
are originally eighteen free parameters in the scalar potential of the 3HDM,
but two are determined by the mass of the $W$ boson and
the mass of the 125 GeV neutral Higgs boson.} free parameters in the 
scalar potential of the 3HDM \cite{Logan}.
There are theoretical constraints on these sixteen
parameters from requiring the stability of the vacuum, 
the absence of charge breaking minima, 
and compliance with unitarity of scattering processes etc.
Such constraints are well-known in the 2HDM 
(e.g. see \cite{Eberhardt:2013uba} for a recent study)
and have been discussed for the scalar potential of the 3HDM 
in \cite{Ivanov:2010wz,Bento:2017eti}. 

In our analysis we do not
impose these constraints because 
they would only rule out certain regions of the parameter space of 
sixteen variables.
As mentioned earlier, the phenomenology
in the charged Higgs sector depends on only five parameters 
(which we take as unconstrained parameters in the above ranges). 
We assume that the freedom in the
remaining eleven parameters can be used to comply with the above 
theoretical constraints while allowing the five parameters in the charged
Higgs sector to vary in the above ranges.
 To justify this 
approach we note that the analogous constraints on the scalar potential in 
2HDMs do not restrict the allowed ranges of the 
two parameters in the charged Higgs sector ($m_{H^\pm}$ and $\tan\beta$) 
due to the freedom in the remaining four 
parameters (for the case of a 2HDM scalar potential 
with only soft breaking terms of a $Z_2$ symmetry).
It is experimental data from processes involving 
$H^\pm$ that constrain the ranges of the 
parameters of the charged Higgs sector
in a 2HDM, and we carry this conclusion   
across to the charged Higgs sector of the 3HDM.

\begin{figure}[phtb]
\includegraphics[width=79mm]{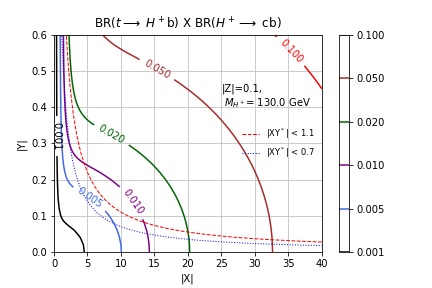}\hspace{4.0mm}
\includegraphics[width=79mm]{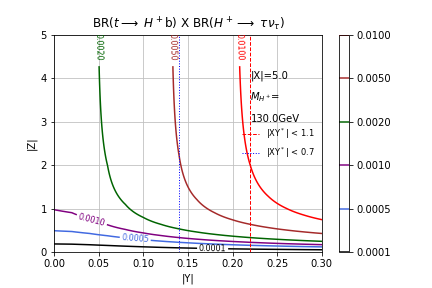}
 \caption{Left panel: Contours of BR$(t\to H^\pm b)\times{\rm BR}(H^\pm\to cb)$
in the plane $[|X|,|Y|]$ with $m_{H^\pm}=130$ GeV and $|Z|=0.1$.
Right panel: Contours of BR$(t\to H^\pm b)\times{\rm BR}(H^\pm\to \tau\nu)$
in the plane $[|Y|,|Z|]$ with $m_{H^\pm}=130$ GeV and $|X|=5$.}
\label{MH130XYplane}
\end{figure} 
In the left panel of Fig.~(\ref{MH130XYplane}) we show contours of BR$(t\to H^\pm b)\times$BR($H^\pm\to cb$)
in the plane $[|X|,|Y|]$ for $|Z|=0.1$ and $m_{H^\pm}=130$ GeV. This is 
an update of a figure
in \cite{Akeroyd2} in which the contours have been chosen to reflect 
the current and
future sensitivity of the LHC. The region consistent with $b\to s\gamma$ lies 
below the
curves of $|XY^*|\le 0.7$ or $|XY^*|\le 1.1$, depending on the sign of 
Re$(XY^*)$ in eq.~(\ref{bsgamma}). In this figure we take $|X|$ and $|Y|$ as 
independent
parameters and thus we do not consider them to be functions of the four 
parameters in 
eq.(\ref{4param}) as in a 3HDM. As mentioned at the end of section II.C, in 
models such as the Aligned 2HDM and a 4HDM the parameters 
$X$, $Y$, and $Z$ would depend on more than four parameters.
The results in the left panel of Fig.~(\ref{MH130XYplane}) are a model independent approach in which the
allowed region of $[|X|,|Y|]$ (for a given $|Z|$) are shown.  
For the chosen value of $m_{H^\pm}=130$ GeV the current limit on BR$(t\to H^\pm b)\times$BR($H^\pm\to cb$) 
is $\le 0.005$ \cite{Sirunyan:2018dvm}. It can be seen from the left panel of Fig.~(\ref{MH130XYplane}) that the current
limit is ruling out parameter space that is permitted by $b\to s\gamma$. The contour with 0.001
will hopefully be approached with 150 fb$^{-1}$ at $\sqrt s=13$ TeV, and such a search would further probe
parameter space of $[|X|,|Y|]$, for a given $|Z|$, that is still allowed by $b\to s\gamma$.

In the right panel of  Fig.~(\ref{MH130XYplane}) we show contours of BR$(t\to H^\pm b)\times$BR($H^\pm\to \tau\nu$)
in the plane $[|Y|,|Z|]$ for $|X|=5$. This is also a model independent approach, and such a plot was not
shown in  \cite{Akeroyd2}. In this case the region allowed by $b\to s\gamma$ 
lies to the left of the perpendicular lines. For the chosen value of $m_{H^\pm}=130$ GeV the current limit on
BR$(t\to H^\pm b)\times$BR($H^\pm\to \tau\nu$) is $\le 0.001$ \cite{Aaboud:2018gjj}. 
It can be seen from the right panel of Fig.~(\ref{MH130XYplane}) that the current
limit is ruling out large regions of parameter space that is permitted by $b\to s\gamma$. The contours with 0.005
and below will hopefully be approached with 150 fb$^{-1}$ at $\sqrt s=13$ TeV, and such a search would further probe
parameter space of $[|Y|,|Z|]$, for a given $|X|$, that is still allowed by $b\to s\gamma$.

We now show our results for the flipped 3HDM and the democratic 3HDM. 
In the other 3HDMs (Type I, Type II and Lepton-specific) 
we have checked that a large BR$(H^\pm\to cb)$ is not 
possible, and the maximum value is typically of the order of a few percent.
\begin{figure}[phtb]
\includegraphics[width=79mm]{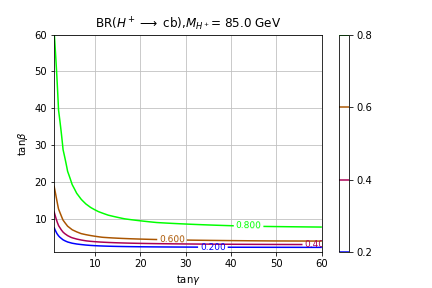}\hspace{4.0mm}
\includegraphics[width=79mm]{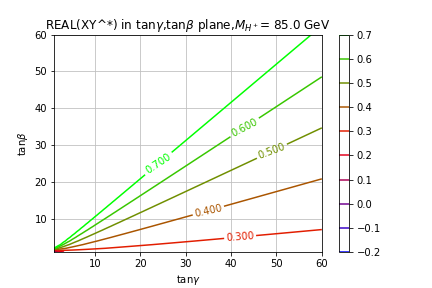}
 \caption{The flipped 3HDM with $\theta=-\pi/3$, $\delta=0$,
and $m_{H^\pm}=85$ GeV.
Left panel: Contours of BR$(H^\pm\to cb)$
in the plane $[\tan\gamma,\tan\beta]$.
Right panel: Contours of Re$(XY^*)$ in the plane $[\tan\gamma,\tan\beta]$.
The allowed parameter space lies below the contour of Re$(XY^*)=0.7$.}
\label{Flipped85cbXY}
\end{figure} 
In Fig.~(\ref{Flipped85cbXY}) we consider the flipped 3HDM with 
$\theta=-\pi/3$, $\delta=0$,
and $m_{H^\pm}=85$ GeV. In the left panel of Fig.~(\ref{Flipped85cbXY}) 
we show contours of BR$(H^\pm\to cb)$ in the plane $[\tan\gamma,\tan\beta]$.
It is evident that for $\tan\gamma\ge 5$ and $\tan\beta\ge 5$ one has
 BR$(H^\pm\to cb)\ge 60\%$, and for $\tan\gamma\ge 10$ and $\tan\beta\ge 10$
the maximum value of around $80\%$ is obtained. However, not all of
this parameter space of $[\tan\gamma,\tan\beta]$ survives the 
constraint from $b\to s\gamma$. This can be seen in the right panel of 
Fig.~(\ref{Flipped85cbXY}) in which
we show contours of Re$(XY^*)$ in the plane $[\tan\gamma,\tan\beta]$.
The allowed parameter space lies below the contour of Re$(XY^*)=0.7$,
and roughly corresponds to the parameter space of $\tan\gamma > \tan\beta$.
By comparing the left and right panels it is clear that a large parameter
space for a dominant BR$(H^\pm\to cb)\ge 60\%$ (left panel) 
survives the $b\to s\gamma$ constraint (right panel). 
Taking a non-zero value of $\delta$ would only lead to slight
modifications of BR$(H^\pm\to cb)$, but would change the regions allowed by
$b\to s\gamma$ due to $X$ and $Y$ both gaining an imaginary part. For
$\delta=0$ the constraint in eq.~(\ref{Imxy}) from the electric dipole moment
of the neutron is automatically satisfied. For $\delta\ne 0$ this latter constraint would rule
out parameter space, and we will consider this scenario later for the democratic 3HDM.

\begin{figure}[phtb]
\includegraphics[width=79mm]{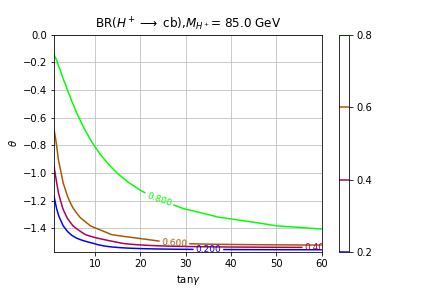}\hspace{4.0mm}
\includegraphics[width=79mm]{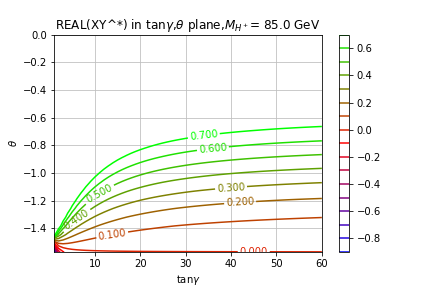}
 \caption{The flipped 3HDM with  $\tan\beta=10$, $\delta=0$, and 
$m_{H^\pm}=85$ GeV.
Left panel: Contours of BR$(H^\pm\to cb)$
in the plane $[\tan\gamma,\theta]$.
Right panel: Contours of Re$(XY^*)$ in the plane $[\tan\gamma,\theta]$.
The allowed parameter space lies below the contour of Re$(XY^*)=0.7$.}
\label{Flippedcb2}
\end{figure}
In Fig.~(\ref{Flippedcb2}) we consider the flipped 3HDM with  
$m_{H^\pm}=85$ GeV but now with 
$\tan\beta=10$ and $\delta=0$. In the left panel of Fig.~(\ref{Flippedcb2}) 
we show contours of BR$(H^\pm\to cb)$
in the plane $[\tan\gamma,\theta]$.
In the right panel of Fig.~(\ref{Flippedcb2})
we show contours of Re$(XY^*)$ in the plane  $[\tan\gamma,\theta]$.
There is a large parameter space
for a dominant BR$(H^\pm\to cb)$ which corresponds to large values
of $\tan\gamma$ and less negative values of $\theta$.
In the right panel of Fig.~(\ref{Flippedcb2}) the  
parameter space allowed by $b\to s\gamma$ 
lies below the contour of Re$(XY^*)=0.7$, and 
thus a large parameter
space for a dominant BR$(H^\pm\to cb)\ge 60\%$ (left panel) 
survives the $b\to s\gamma$ constraint (right panel). In summary, from the results
in Fig.~(\ref{Flipped85cbXY}) and Fig.~(\ref{Flippedcb2}) it is clear
that a large part of the $[\tan\gamma$, $\tan\beta$, $\theta]$ parameter space
(with $\delta=0$)
gives rise to a dominant BR$(H^\pm\to cb)$ while complying with 
constraints from $b\to s\gamma$. As mentioned earlier, we consider 
the right panels of Fig.~(\ref{Flipped85cbXY}) and Fig.~(\ref{Flippedcb2})
to be representative of the true constraints on the planes  
$[\tan\gamma$, $\tan\beta]$ and $[\tan\gamma,\theta]$ from $b\to s\gamma$.
We expect that the true excluded region would be shifted somewhat from 
the excluded regions in Fig.~(\ref{Flipped85cbXY}) and Fig.~(\ref{Flippedcb2}),
but it would not increase significantly in area. 
Given the large parameter space for a dominant 
BR$(H^\pm\to cb)$ in the flipped 3HDM we expect a sizeable region 
of large BR to survive. Taking a non-zero value of $\delta$ would only lead to slight
modifications of the above plots for BR$(H^\pm\to cb)$, but would have an effect
on the plot for Re$(XY^*)$. We will illustrate this when we consider the democratic 3HDM below.

\begin{figure}[phtb]
\includegraphics[width=79mm]{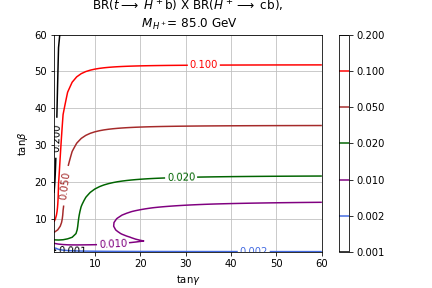}\hspace{4.0mm}
\includegraphics[width=79mm]{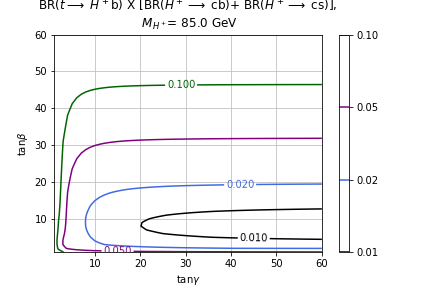}
 \caption{The flipped 3HDM with  $\theta=-\pi/3$, $\delta=0$, and 
$m_{H^\pm}=85$ GeV.
Left panel: Contours of BR$(t\to H^\pm b)\times{\rm BR}(H^\pm\to cb)$
in the plane $[\tan\gamma,\tan\beta]$.
Right panel: Contours of BR$(t\to H^\pm b)\times [{\rm BR}(H^\pm\to cb)$+
BR$(H^\pm\to cs)]$ 
in the plane $[\tan\gamma,\tan\beta]$.}
\label{Flipped85tbcbcs}
\end{figure}

In Fig.~(\ref{Flipped85tbcbcs}) we take the input parameters
of Fig.~(\ref{Flipped85cbXY}) for the flipped 3HDM. In the left panel we 
plot contours of 
BR$(t\to H^\pm b)\times {\rm BR}(H^\pm\to cb$) in the plane 
$[\tan\beta, \tan\gamma]$. This is the 
product that is being constrained by the CMS search
at the LHC using three $b-$tags \cite{Sirunyan:2018dvm}.
However, for $m_{H^\pm}=85$ GeV (which is used in the 
Fig.~(\ref{Flipped85tbcbcs})) there is no limit on 
BR$(t\to H^\pm b)\times {\rm BR}(H^\pm\to cb$) from the LHC. The
only limit is $\le 20\%$ from the Tevatron \cite{Abazov:2009aa},
using a strategy that was sensitive to any quark decay mode of $H^\pm$.
We plot contours of BR$(t\to H^\pm b)\times {\rm BR}(H^\pm\to cb$)
with values of $0.2$ to $0.002$. The region of the $[\tan\beta, \tan\gamma]$
plane that is above the contour of $0.2$ is ruled out, while 
the region below corresponds to a potential discovery of such an $H^\pm$. 
It is hoped that future searches of the LHC with $\sqrt s=13$ TeV and 
150 fb$^{-1}$ (or more) of data will have sensitivity to 
BR$(t\to H^\pm b)\times {\rm BR}(H^\pm\to cb$) of 0.02 or below.

In the right panel of Fig.~(\ref{Flipped85tbcbcs}) contours of 
BR$(t\to H^\pm b)\times {\rm BR}[(H^\pm\to cb)+{\rm BR}(H^\pm\to cs)]$
are plotted with $m_{H^\pm}=85$ GeV. This 
product is the observable that is being constrained by the
searches that use $2b$ tags \cite{Aad:2013hla,Khachatryan:2015uua}, and 
the figure is very similar to the left panel of  Fig.~(\ref{Flipped85tbcbcs}). 
However, to obtain sensitivity to a given contour 
we expect that the $2b$ search will require more integrated
luminosity than the $3b$ search, because the latter has smaller backgrounds
as discussed earlier in section III.B.2.

\begin{figure}[phtb]
\includegraphics[width=79mm]{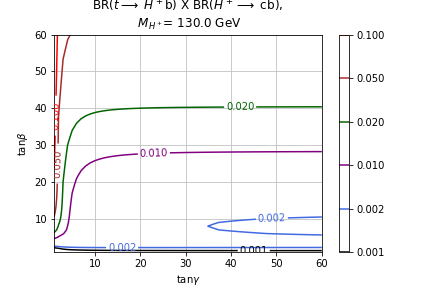}\hspace{4.0mm}
\includegraphics[width=79mm]{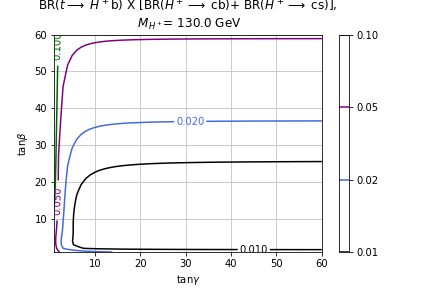}
 \caption{The flipped 3HDM with  $\theta=-\pi/3$, $\delta=0$, and 
$m_{H^\pm}=130$ GeV.
Left panel: Contours of BR$(t\to H^\pm b)\times{\rm BR}(H^\pm\to cb)$
in the plane $[\tan\gamma,\tan\beta]$.
Right panel: Contours of BR$(t\to H^\pm b)\times [{\rm BR}(H^\pm\to cb)$+
BR$(H^\pm\to cs)]$ 
in the plane $[\tan\gamma,\tan\beta]$.}
\label{Flipped130tbcbcs}
\end{figure}
Fig.~(\ref{Flipped130tbcbcs}) is the same as Fig.~(\ref{Flipped85tbcbcs}) but with $m_{H^\pm}=130$ GeV, and hence 
BR($t\to H^\pm b$) is reduced compared to the corresponding case with $m_{H^\pm}=85$ GeV. However,
in both panels in Fig.~(\ref{Flipped130tbcbcs}) the current excluded region is roughly above the contour of 0.02 (instead of 0.2)
due to the LHC searches \cite{Sirunyan:2018dvm,Aad:2013hla,Khachatryan:2015uua} having superior sensitivity to those of the Tevatron in
the region 90 GeV$\le m_{H^\pm}\le 160$ GeV.
It can be seen that a sizeable area of the  $[\tan\gamma,\tan\beta]$ parameter space is ruled out, while the
region below the 0.02 contour would provide a possible signal for $H^\pm$. It is hoped that future searches will
have sensitivity to contours of 0.001 in both the $2b$ and $3b$ channels for 90 GeV$\le m_{H^\pm}\le 160$ GeV.

\begin{figure}[phtb]
\includegraphics[width=79mm]{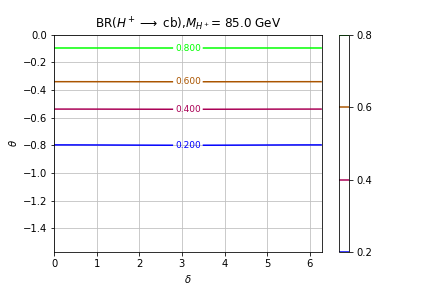}\hspace{4.0mm}
\includegraphics[width=79mm]{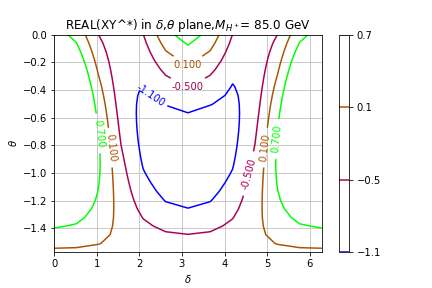}
 \caption{The democratic 3HDM with $\tan\beta=40$, $\tan\gamma=10$,
and $m_{H^\pm}=85$ GeV.
Left panel: Contours of BR$(H^\pm\to cb)$
in the plane $[\delta,\theta]$.
Right panel: Contours of Re$(XY^*)$ in the plane $[\delta,\theta]$.
The allowed parameter space lies within the range $-1.1\le
 {\rm Re}(XY^*)\le 0.7$.}
\label{DemoreXY}
\end{figure} 

\begin{figure}[phtb]
\includegraphics[width=79mm]{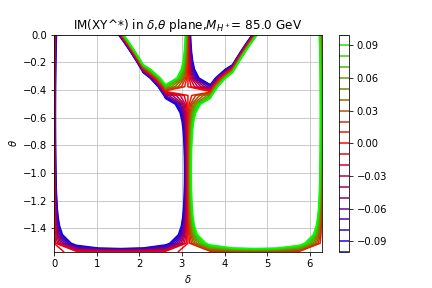}
 \caption{The democratic 3HDM with $\tan\beta=40$, $\tan\gamma=10$, 
and $m_{H^\pm}=85$ GeV. Contours of Im$(XY^*)$ in the plane $[\delta,\theta]$.
The allowed parameter space lies within the range $|{\rm Im}(XY^*)|=0.1$.}
\label{DemoimXY}
\end{figure} 

We now show results in the democratic 3HDM. Taking $\delta=0$ we find that large values of BR$(H^\pm\to cb)$ are 
possible in regions
of parameter space, but these regions are essentially 
ruled out by the $b\to s\gamma$ constraint. However, by 
taking $\delta\ne 0$ there are regions that have a large BR$(H^\pm\to cb)$ 
while complying with the constraints
from $b\to s\gamma$ and the electric dipole moment of the neutron. 
In Fig.~(\ref{DemoreXY}) we take $\tan\beta=40$, $\tan\gamma=10$,
and $m_{H^\pm}=85$ GeV in the democratic 3HDM. 
In the left panel, contours of BR$(H^\pm\to cb)$ are plotted 
in the plane $[\delta,\theta]$. It can be seen that large values
of  BR$(H^\pm\to cb)$ are possible, but $\delta$ has almost
no effect on its magnitude. In the right panel of Fig.~(\ref{DemoreXY}) 
we plot contours of Re$(XY^*)$ in the plane $[\delta,\theta]$, 
and the allowed parameter space lies within the range $-1.1\le
{\rm Re}(XY^*)\le 0.7$. One can see that varying $\delta$ has a sizeable
effect on ${\rm Re}(XY^*)$. By comparing the left and right panels 
it can be seen that  
the region $1\le \delta \le 5$ and $0\ge \theta\ge -0.5$ 
gives a large BR$(H^\pm\to cb)$ that is also compatible with the $b\to s\gamma$
constraint. However, this region is further constrained by 
Fig.~(\ref{DemoimXY}) in which we plot contours of Im$(XY^*)$ in the 
plane $[\delta,\theta]$, and the allowed parameter space lies within the 
range $|{\rm Im}(XY^*)|\le 0.1$. There are three allowed strips (with
one being around $\delta=\pi$) in the region of
large BR$(H^\pm\to cb)$ (i.e. $0\ge \theta\ge -0.5$). Consequently, 
the democratic 3HDM is a candidate model for a possible signal in
future $3b$ searches for $H^\pm$ as carried out in \cite{Sirunyan:2018dvm}, although
the parameter space for a large  BR$(H^\pm\to cb)$ is much smaller than that
in the flipped 3HDM, and is likely to require $\delta\ne 0$.

\begin{figure}[phtb]
\includegraphics[width=79mm]{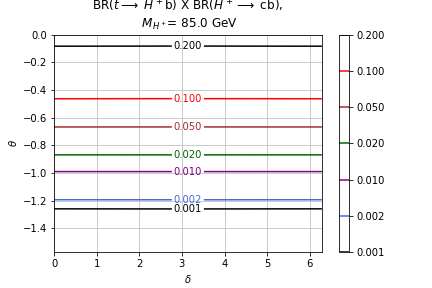}\hspace{4.0mm}
\includegraphics[width=79mm]{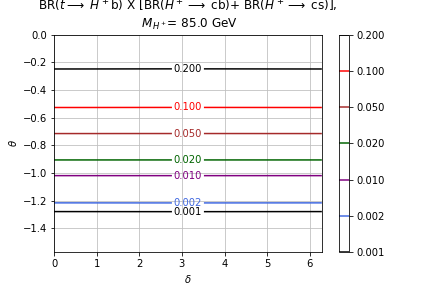}
 \caption{The democratic 3HDM with $\tan\beta=40$, $\tan\gamma=10$, 
and $m_{H^\pm}=85$ GeV.
Left panel: Contours of BR$(t\to H^\pm b)\times{\rm BR}(H^\pm\to cb)$
in the plane $[\theta,\delta]$.
Right panel: Contours of BR$(t\to H^\pm b)\times [{\rm BR}(H^\pm\to cb)$+
BR$(H^\pm\to cs)]$ 
in the plane $[\theta,\delta]$.}
\label{Demo85tbcbcs}
\end{figure}

\begin{figure}[phtb]
\includegraphics[width=79mm]{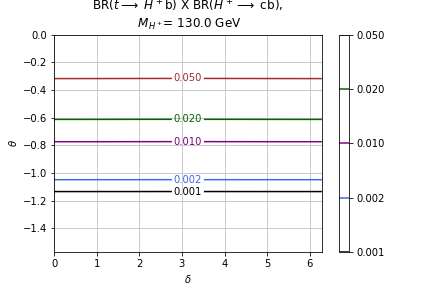}\hspace{4.0mm}
\includegraphics[width=79mm]{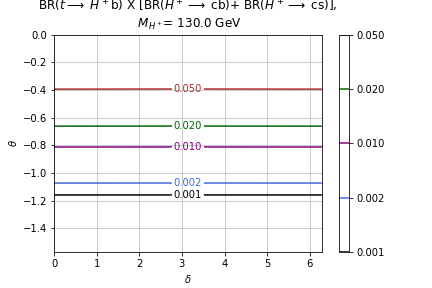}
 \caption{The democratic 3HDM with $\tan\beta=40$, $\tan\gamma=10$, 
and $m_{H^\pm}=130$ GeV.
Left panel: Contours of BR$(t\to H^\pm b)\times{\rm BR}(H^\pm\to cb)$
in the plane $[\theta,\delta]$.
Right panel: Contours of BR$(t\to H^\pm b)\times [{\rm BR}(H^\pm\to cb)$+
BR$(H^\pm\to cs)]$ 
in the plane $[\theta,\delta]$.}
\label{Demo130tbcbcs}
\end{figure}

Fig.~(\ref{Demo85tbcbcs}) and Fig.~(\ref{Demo130tbcbcs}) are with the same parameter
choice of $\tan\beta=40$, $\tan\gamma=10$ in the democratic 3HDM, and are 
the plots that correspond to Fig.~(\ref{Flipped85tbcbcs}) and 
Fig.~(\ref{Flipped130tbcbcs}) in the flipped 3HDM. 
Contours of BR$(t\to H^\pm b)\times{\rm BR}(H^\pm\to cb)$ and 
BR$(t\to H^\pm b)\times [{\rm BR}(H^\pm\to cb)$+BR$(H^\pm\to cs)]$ are plotted in the plane $[\delta,\theta]$.
The (small) allowed region can be read off from Fig.~(\ref{DemoimXY}), most of it
being around $\delta=\pi$, but with $-0.6\le \theta\le -1.2$ being excluded from Fig.~(\ref{DemoreXY}).

\section{Conclusions}  
In summary, we have studied a 3HDM wherein two charged Higgs bosons states exist, one of which we have assumed to be lighter than the top quark and the other one heavier. Hence, the light state can be produced in (anti)top decays via
$t\to bH^\pm$, particularly at hadron colliders like the LHC, via the 
$pp\to t\bar t$ process, which herein has a significant cross section 
(nearing the nb level) so that the main focus of our analysis has been on this $H^\pm$ production channel. Amongst the possible $H^\pm$ decay modes in a 3HDM we have selected here the fermionic ones, i.e. $H^\pm \to cs, cb,$ and $\tau\nu$, which are those exploited in collider searches, at both past (LEP and Tevatron)  and present (LHC) machines. Amongst these three channels, we have concentrated on $H^\pm\to cb$ as it offers a twofold experimental advantage. On the one hand, the irreducible background from $W^\pm\to cb$ decays is suppressed by the  Cabibbo-Kobayashi-Maskawa (CKM) matrix.  On the other hand, it can be filtered out by requiring a $b$--tag of one the two jets that 
eventually emerges in the detector. Furthermore, from a theoretical point of view 
this decay mode may be a privileged probe of the underlying 3HDM structure. This is 
because the BR$(H^\pm \to cb)$ can be large in the flipped and democratic versions of 
the 3HDM, but not in the  type I, type II, and 
lepton-specific structures, while being compatible with experimental constraints, chiefly, those from $b\to s \gamma$. 

We have then performed the first comprehensive study of the decay mode $H^\pm \to cb$ 
in terms of the four fundamental parameters of the charged Higgs sector of the
3HDM ($\beta,\gamma,\theta,$ and $\delta$) over the available $m_{H^\pm}$ range.
We found that the parameter space for a large BR$(H^\pm \to cb)$ is much bigger 
in the flipped 3HDM than in the democratic 3HDM.
Our emphasis has been on the interval $80$ GeV $< m_{H^\pm} < 90$ GeV, to which the LHC
has no sensitivity at present, the reason being that no experimental searches have 
yet been attempted for the decays $H^\pm\to cb/cs$ at this collider.
For the purpose of encouraging such searches, 
we have mapped out the 3HDM parameter spaces of the flipped and democratic types that 
can be accessible at the LHC as a function of its increased luminosity, 
concluding that they should be accessible in the near future 
by exploiting established experimental techniques. In fact, this can be achieved by 
resorting to both appearance and disappearance searches. The former would have direct 
sensitivity to the $H^\pm\to cb$ channel while the latter would have indirect 
access one to it, via the absence of the expected number of 
$W^\pm\to \ell\nu$ ($\ell=e,\mu$) and $W\to q\overline q$ events originating
from $pp\to t\overline t$ with standard top decay for both $t$ and $\overline t$. 
Similarly positive prospects are expected for future $e^+e^-$ colliders,  
like FCC-ee, CEPC, and ILC, where the $H^\pm$ state would be pair produced via 
$e^+e^-\to H^+H^-$. At such colliders the QCD backgrounds are much reduced
with respect to those at the LHC, which greatly facilitates the extraction of 
the $H^\pm\to cb$ mode.

\section*{Acknowledgements}
SM is funded in part through the NExT Institute, the STFC CG ST/L000296/1 and the H2020-MSCA-RISE-2014 grant no.  645722
 (NonMinimalHiggs).


\begin{thebibliography}{99}

%\cite{Aad:2012tfa}
\bibitem{Aad:2012tfa} 
  G.~Aad {\it et al.} [ATLAS Collaboration],
  %``Observation of a new particle in the search for the Standard Model Higgs boson with the ATLAS detector at the LHC,''
  Phys.\ Lett.\ B {\bf 716}, 1 (2012).
%  doi:10.1016/j.physletb.2012.08.020
%  [arXiv:1207.7214 [hep-ex]].
  %%CITATION = doi:10.1016/j.physletb.2012.08.020;%%
  %6047 citations counted in INSPIRE as of 17 May 2016

%\cite{Chatrchyan:2012xdj}
\bibitem{Chatrchyan:2012xdj} 
  S.~Chatrchyan {\it et al.} [CMS Collaboration],
  %``Observation of a new boson at a mass of 125 GeV with the CMS experiment at the LHC,''
  Phys.\ Lett.\ B {\bf 716}, 30 (2012).
%  doi:10.1016/j.physletb.2012.08.021
%  [arXiv:1207.7235 [hep-ex]].
  %%CITATION = doi:10.1016/j.physletb.2012.08.021;%%
  %5904 citations counted in INSPIRE as of 17 May 2016

%\cite{Aaboud:2018zhk}
\bibitem{Aaboud:2018zhk} 
  M.~Aaboud {\it et al.} [ATLAS Collaboration],
  %``Observation of $H \rightarrow b\bar{b}$ decays and $VH$ production with the ATLAS detector,''
  arXiv:1808.08238 [hep-ex].
  %%CITATION = ARXIV:1808.08238;%%

%\cite{Branco:2011iw}
\bibitem{Branco:2011iw} 
  G.~C.~Branco, P.~M.~Ferreira, L.~Lavoura, M.~N.~Rebelo, M.~Sher and J.~P.~Silva,
  %``Theory and phenomenology of two-Higgs-doublet models,''
  Phys.\ Rept.\  {\bf 516}, 1 (2012)
%  doi:10.1016/j.physrep.2012.02.002
  [arXiv:1106.0034 [hep-ph]].
  %%CITATION = doi:10.1016/j.physrep.2012.02.002;%%
  %1292 citations counted in INSPIRE as of 29 Aug 2018

%\cite{Akeroyd:2016ymd}
\bibitem{Akeroyd:2016ymd} 
  A.~G.~Akeroyd {\it et al.},
  %``Prospects for charged Higgs searches at the LHC,''
  Eur.\ Phys.\ J.\ C {\bf 77}, no. 5, 276 (2017)
%  doi:10.1140/epjc/s10052-017-4829-2
  [arXiv:1607.01320 [hep-ph]].
  %%CITATION = doi:10.1140/epjc/s10052-017-4829-2;%%
  %48 citations counted in INSPIRE as of 29 Aug 2018


\bibitem{Grossman} 
  Y.~Grossman,
  %``Phenomenology of models with more than two Higgs doublets,''
  Nucl.\ Phys.\ B {\bf 426}, 355 (1994).
 % [hep-ph/9401311].
  %%CITATION = doi:10.1016/0550-3213(94)90316-6;%%
  %238 citations counted in INSPIRE as of 10 Feb 2016

%\cite{Akeroyd:1994ga}
\bibitem{Akeroyd:1994ga} 
  A.~G.~Akeroyd and W.~J.~Stirling,
  %``Light charged Higgs scalars at high-energy e+ e- colliders,''
  Nucl.\ Phys.\ B {\bf 447}, 3 (1995).
%  doi:10.1016/0550-3213(95)00173-P
  %%CITATION = doi:10.1016/0550-3213(95)00173-P;%%
  %50 citations counted in INSPIRE as of 07 May 2016

%\cite{Akeroyd:1995cf}
\bibitem{Akeroyd:1995cf} 
  A.~G.~Akeroyd,
  %``Hidden top quark decays to charged Higgs scalars at the Tevatron,''
  hep-ph/9509203.
  %%CITATION = HEP-PH/9509203;%%
  %12 citations counted in INSPIRE as of 17 May 2016


%\cite{Akeroyd2}
\bibitem{Akeroyd2} 
  A.~G.~Akeroyd, S.~Moretti and J.~Hernandez-Sanchez,
  %``Light charged Higgs bosons decaying to charm and bottom quarks in models with two or more Higgs doublets,''
  Phys.\ Rev.\ D {\bf 85}, 115002 (2012).
%  doi:10.1103/PhysRevD.85.115002
%  [arXiv:1203.5769 [hep-ph]].
  %%CITATION = doi:10.1103/PhysRevD.85.115002;%%
  %13 citations counted in INSPIRE as of 17 May 2016

%\cite{Akeroyd:2016ssd}
\bibitem{Akeroyd:2016ssd} 
  A.~G.~Akeroyd, S.~Moretti, K.~Yagyu and E.~Yildirim,
  %``Light charged Higgs boson scenario in 3-Higgs doublet models,''
  Int.\ J.\ Mod.\ Phys.\ A {\bf 32}, no. 23n24, 1750145 (2017)
%  doi:10.1142/S0217751X17501457
  [arXiv:1605.05881 [hep-ph]].
  %%CITATION = doi:10.1142/S0217751X17501457;%%
  %6 citations counted in INSPIRE as of 22 Jun 2018

 
%\cite{Glashow:1976nt}
\bibitem{Glashow:1976nt} 
  S.~L.~Glashow and S.~Weinberg,
  %``Natural Conservation Laws for Neutral Currents,''
  Phys.\ Rev.\ D {\bf 15}, 1958 (1977);
%  doi:10.1103/PhysRevD.15.1958
  %%CITATION = doi:10.1103/PhysRevD.15.1958;%%
  %1650 citations counted in INSPIRE as of 06 Sep 2018
%\cite{Paschos:1976ay}
%\bibitem{Paschos:1976ay} 
  E.~A.~Paschos,
  %``Diagonal Neutral Currents,''
  Phys.\ Rev.\ D {\bf 15}, 1966 (1977).
%  doi:10.1103/PhysRevD.15.1966
  %%CITATION = doi:10.1103/PhysRevD.15.1966;%%
  %481 citations counted in INSPIRE as of 06 Sep 2018


%\cite{Sirunyan:2018dvm}
\bibitem{Sirunyan:2018dvm} 
  A.~M.~Sirunyan {\it et al.} [CMS Collaboration],
  %``Search for a charged Higgs boson decaying to charm and bottom quarks in proton-proton collisions at $\sqrt{s} =$ 8 TeV,''
  arXiv:1808.06575 [hep-ex].
  %%CITATION = ARXIV:1808.06575;%%

\bibitem{Barger} 
  V.~D.~Barger, J.~L.~Hewett and R.~J.~N.~Phillips,
  %``New Constraints on the Charged Higgs Sector in Two Higgs Doublet Models,''
  Phys.\ Rev.\ D {\bf 41}, 3421 (1990).
  %%CITATION = doi:10.1103/PhysRevD.41.3421;%%
  %445 citations counted in INSPIRE as of 28 Jan 2016

\bibitem{Logan} 
  G.~Cree and H.~E.~Logan,
  %``Yukawa alignment from natural flavor conservation,''
  Phys.\ Rev.\ D {\bf 84}, 055021 (2011).
  %[arXiv:1106.4039 [hep-ph]].
  %%CITATION = doi:10.1103/PhysRevD.84.055021;%%
  %13 citations counted in INSPIRE as of 25 Jan 2016

%\cite{Jung:2010ik} 
\bibitem{Jung:2010ik} 
  M.~Jung, A.~Pich and P.~Tuzon,
  %``Charged-Higgs phenomenology in the Aligned two-Higgs-doublet model,''
  JHEP {\bf 1011}, 003 (2010)
%  doi:10.1007/JHEP11(2010)003
  [arXiv:1006.0470 [hep-ph]].
  %%CITATION = doi:10.1007/JHEP11(2010)003;%%
  %136 citations counted in INSPIRE as of 29 Aug 2018

%\cite{Trott:2010iz}
\bibitem{Trott:2010iz} 
  M.~Trott and M.~B.~Wise,
  %``On Theories of Enhanced {CP} Violation in $B_{s,d}$ Meson Mixing,''
  JHEP {\bf 1011}, 157 (2010)
%  doi:10.1007/JHEP11(2010)157
  [arXiv:1009.2813 [hep-ph]].
  %%CITATION = doi:10.1007/JHEP11(2010)157;%%
  %33 citations counted in INSPIRE as of 29 Aug 2018


\bibitem{Ciuchini1}
  M.~Ciuchini, E.~Franco, G.~Martinelli, L.~Reina and L.~Silvestrini,
  %``b ---> s gamma and b ---> s g: A Theoretical reappraisal,''
  Phys.\ Lett.\ B {\bf 334}, 137 (1994).
 % [hep-ph/9406239]. 
  %%CITATION = HEP-PH/9406239;%%

\bibitem{Ciuchini2}
  M.~Ciuchini, G.~Degrassi, P.~Gambino and G.~F.~Giudice,
  %``Next-to-leading QCD corrections to B ---> X(s) gamma: Standard model and two Higgs doublet model,''
  Nucl.\ Phys.\ B {\bf 527}, 21 (1998).
  %[hep-ph/9710335].
  %%CITATION = HEP-PH/9710335;%%

\bibitem{Borzumati} 
  F.~Borzumati and C.~Greub,
  %``2HDMs predictions for anti-B ---> X(s) gamma in NLO QCD,''  
Phys.\ Rev.\ D {\bf 58}, 074004 (1998).  
%doi:10.1103/PhysRevD.58.074004  
%[hep-ph/9802391].  
%%CITATION = doi:10.1103/PhysRevD.58.074004;%%  
%303 citations counted in INSPIRE as of 05 Feb 2016

\bibitem{Gambino} 
P.~Gambino and M.~Misiak,
  %``Quark mass effects in anti-B ---> X(s gamma),''
  Nucl.\ Phys.\ B {\bf 611}, 338 (2001).
  %[hep-ph/0104034].
  %%CITATION = HEP-PH/0104034;%%

\bibitem{Misiak}
  T.~Hermann, M.~Misiak and M.~Steinhauser,
  %``$\bar{B}\to X_s \gamma$ in the Two Higgs Doublet Model up to Next-to-Next-to-Leading Order in QCD,''
  JHEP {\bf 1211}, 036 (2012).
%  [arXiv:1208.2788 [hep-ph]].
  %%CITATION = ARXIV:1208.2788;%%

\bibitem{Misiak2}
  M.~Misiak, H.~M.~Asatrian, R.~Boughezal, M.~Czakon, T.~Ewerth, A.~Ferroglia, P.~Fiedler and P.~Gambino {\it et al.},
  %``Updated NNLO QCD predictions for the weak radiative B-meson decays,''
  Phys.\ Rev.\ Lett.\  {\bf 114},  221801 (2015).
  %[arXiv:1503.01789 [hep-ph]].
  %%CITATION = ARXIV:1503.01789;%%
  %15 citations counted in INSPIRE as of 26 juin 2015

%\cite{Moretti:1994ds}
\bibitem{Moretti:1994ds} 
  S.~Moretti and W.~J.~Stirling,
  %``Contributions of below threshold decays to MSSM Higgs branching ratios,''
  Phys.\ Lett.\ B {\bf 347}, 291 (1995)
  Erratum: [Phys.\ Lett.\ B {\bf 366}, 451 (1996)]
%  doi:10.1016/0370-2693(95)00088-3, 10.1016/0370-2693(95)01477-2
  [hep-ph/9412209, hep-ph/9511351].
  %%CITATION = doi:10.1016/0370-2693(95)00088-3, 10.1016/0370-2693(95)01477-2;%%
  %95 citations counted in INSPIRE as of 30 Aug 2018


%\cite{Djouadi:1995gv}
\bibitem{Djouadi:1995gv} 
  A.~Djouadi, J.~Kalinowski and P.~M.~Zerwas,
  %``Two and three-body decay modes of SUSY Higgs particles,''
  Z.\ Phys.\ C {\bf 70}, 435 (1996)
%  doi:10.1007/s002880050121
  [hep-ph/9511342].
  %%CITATION = doi:10.1007/s002880050121;%%
  %191 citations counted in INSPIRE as of 30 Aug 2018

%\cite{Akeroyd:1998dt}
\bibitem{Akeroyd:1998dt} 
  A.~G.~Akeroyd,
  %``Three body decays of Higgs bosons at LEP-2 and application to a hidden fermiophobic Higgs,''
  Nucl.\ Phys.\ B {\bf 544}, 557 (1999).
%  doi:10.1016/S0550-3213(98)00845-1
%  [hep-ph/9806337].
  %%CITATION = doi:10.1016/S0550-3213(98)00845-1;%%
  %49 citations counted in INSPIRE as of 17 May 2016

%\cite{Kling:2015uba}
\bibitem{Kling:2015uba} 
  F.~Kling, A.~Pyarelal and S.~Su,
  %``Light Charged Higgs Bosons to AW/HW via Top Decay,''
  JHEP {\bf 1511}, 051 (2015)
%  doi:10.1007/JHEP11(2015)051
  [arXiv:1504.06624 [hep-ph]].
  %%CITATION = doi:10.1007/JHEP11(2015)051;%%
  %26 citations counted in INSPIRE as of 30 Aug 2018

%\cite{Arhrib:2016wpw}
\bibitem{Arhrib:2016wpw} 
  A.~Arhrib, R.~Benbrik and S.~Moretti,
  %``Bosonic Decays of Charged Higgs Bosons in a 2HDM Type-I,''
  Eur.\ Phys.\ J.\ C {\bf 77}, no. 9, 621 (2017)
%  doi:10.1140/epjc/s10052-017-5197-7
  [arXiv:1607.02402 [hep-ph]].
  %%CITATION = doi:10.1140/epjc/s10052-017-5197-7;%%
  %14 citations counted in INSPIRE as of 30 Aug 2018

%\cite{Arbey:2017gmh}
\bibitem{Arbey:2017gmh} 
  A.~Arbey, F.~Mahmoudi, O.~Stal and T.~Stefaniak,
  %``Status of the Charged Higgs Boson in Two Higgs Doublet Models,''
  Eur.\ Phys.\ J.\ C {\bf 78}, no. 3, 182 (2018)
%  doi:10.1140/epjc/s10052-018-5651-1
  [arXiv:1706.07414 [hep-ph]].
  %%CITATION = doi:10.1140/epjc/s10052-018-5651-1;%%
  %28 citations counted in INSPIRE as of 30 Aug 2018

%\cite{Arhrib:2017wmo}
\bibitem{Arhrib:2017wmo} 
  A.~Arhrib, R.~Benbrik, R.~Enberg, W.~Klemm, S.~Moretti and S.~Munir,
  %``Identifying a light charged Higgs boson at the LHC Run II,''
  Phys.\ Lett.\ B {\bf 774}, 591 (2017)
%  doi:10.1016/j.physletb.2017.10.006
  [arXiv:1706.01964 [hep-ph]].
  %%CITATION = doi:10.1016/j.physletb.2017.10.006;%%
  %4 citations counted in INSPIRE as of 30 Aug 2018


%\cite{Aoki:2009ha}
\bibitem{Aoki:2009ha} 
  M.~Aoki, S.~Kanemura, K.~Tsumura and K.~Yagyu,
  %``Models of Yukawa interaction in the two Higgs doublet model, and their collider phenomenology,''
  Phys.\ Rev.\ D {\bf 80}, 015017 (2009)
%  doi:10.1103/PhysRevD.80.015017
  [arXiv:0902.4665 [hep-ph]].
  %%CITATION = doi:10.1103/PhysRevD.80.015017;%%
  %240 citations counted in INSPIRE as of 06 Sep 2018

%\cite{Logan:2010ag}
\bibitem{Logan:2010ag} 
  H.~E.~Logan and D.~MacLennan,
  %``Charged Higgs phenomenology in the flipped two Higgs doublet model,''
  Phys.\ Rev.\ D {\bf 81}, 075016 (2010)
%  doi:10.1103/PhysRevD.81.075016
  [arXiv:1002.4916 [hep-ph]].
  %%CITATION = doi:10.1103/PhysRevD.81.075016;%%
  %41 citations counted in INSPIRE as of 06 Sep 2018

%\cite{Pich:2009sp}
\bibitem{Pich:2009sp} 
  A.~Pich and P.~Tuzon,
  %``Yukawa Alignment in the Two-Higgs-Doublet Model,''
  Phys.\ Rev.\ D {\bf 80}, 091702 (2009)
%  doi:10.1103/PhysRevD.80.091702
  [arXiv:0908.1554 [hep-ph]].
  %%CITATION = doi:10.1103/PhysRevD.80.091702;%%
  %246 citations counted in INSPIRE as of 06 Sep 2018

%\cite{HernandezSanchez:2012eg}
\bibitem{HernandezSanchez:2012eg} 
  J.~Hernandez-Sanchez, S.~Moretti, R.~Noriega-Papaqui and A.~Rosado,
  %``Off-diagonal terms in Yukawa textures of the Type-III 2-Higgs doublet model and light charged Higgs boson phenomenology,''
  JHEP {\bf 1307}, 044 (2013).
%  doi:10.1007/JHEP07(2013)044
%  [arXiv:1212.6818 [hep-ph]].
  %%CITATION = doi:10.1007/JHEP07(2013)044;%%
  %25 citations counted in INSPIRE as of 01 Oct 2018


%\cite{Abazov:2009aa}
\bibitem{Abazov:2009aa} 
  V.~M.~Abazov {\it et al.} [D0 Collaboration],
  %``Search for Charged Higgs Bosons in Top Quark Decays,''
  Phys.\ Lett.\ B {\bf 682}, 278 (2009)
%  doi:10.1016/j.physletb.2009.11.016
  [arXiv:0908.1811 [hep-ex]].
  %%CITATION = doi:10.1016/j.physletb.2009.11.016;%%
  %162 citations counted in INSPIRE as of 20 Jun 2018

%\cite{Aaltonen:2009ke}
\bibitem{Aaltonen:2009ke} 
  T.~Aaltonen {\it et al.} [CDF Collaboration],
  %``Search for charged Higgs bosons in decays of top quarks in p anti-p collisions at s**(1/2) = 1.96 TeV,''
  Phys.\ Rev.\ Lett.\  {\bf 103}, 101803 (2009)
%  doi:10.1103/PhysRevLett.103.101803
  [arXiv:0907.1269 [hep-ex]].
  %%CITATION = doi:10.1103/PhysRevLett.103.101803;%%
  %144 citations counted in INSPIRE as of 20 Jun 2018

 


%\cite{Chatrchyan:2012vca}
\bibitem{Chatrchyan:2012vca} 
  S.~Chatrchyan {\it et al.} [CMS Collaboration],
  %``Search for a light charged Higgs boson in top quark decays in $pp$ collisions at $\sqrt{s}=7$ TeV,''
  JHEP {\bf 1207}, 143 (2012)
%  doi:10.1007/JHEP07(2012)143
  [arXiv:1205.5736 [hep-ex]].
  %%CITATION = doi:10.1007/JHEP07(2012)143;%%
  %213 citations counted in INSPIRE as of 29 Jun 2018

%\cite{Aad:2012rjx}
\bibitem{Aad:2012rjx} 
  G.~Aad {\it et al.} [ATLAS Collaboration],
  %``Search for charged Higgs bosons through the violation of lepton universality in $t\bar{t}$ events using $pp$ collision data at $\sqrt{s}=7$ TeV with the ATLAS experiment,''
  JHEP {\bf 1303}, 076 (2013)
%  doi:10.1007/JHEP03(2013)076
  [arXiv:1212.3572 [hep-ex]].
  %%CITATION = doi:10.1007/JHEP03(2013)076;%%
  %60 citations counted in INSPIRE as of 29 Jun 2018

%\cite{Aad:2012tj}
\bibitem{Aad:2012tj} 
  G.~Aad {\it et al.} [ATLAS Collaboration],
  %``Search for charged Higgs bosons decaying via $H^{+} \to \tau \nu$ in top quark pair events using $pp$ collision data at $\sqrt{s}=7$ TeV with the ATLAS detector,''
  JHEP {\bf 1206}, 039 (2012)
%  doi:10.1007/JHEP06(2012)039
  [arXiv:1204.2760 [hep-ex]].
  %%CITATION = doi:10.1007/JHEP06(2012)039;%%
  %230 citations counted in INSPIRE as of 03 Jul 2018


%\cite{Aad:2014kga}
\bibitem{Aad:2014kga} 
  G.~Aad {\it et al.} [ATLAS Collaboration],
  %``Search for charged Higgs bosons decaying via $H^{\pm} \rightarrow \tau^{\pm}\nu$ in fully hadronic final states using $pp$ collision data at $\sqrt{s} = 8$ TeV with the ATLAS detector,''
  JHEP {\bf 1503}, 088 (2015)
%  doi:10.1007/JHEP03(2015)088
  [arXiv:1412.6663 [hep-ex]].
  %%CITATION = doi:10.1007/JHEP03(2015)088;%%
  %158 citations counted in INSPIRE as of 29 Jun 2018

%\cite{Khachatryan:2015qxa}
\bibitem{Khachatryan:2015qxa} 
  V.~Khachatryan {\it et al.} [CMS Collaboration],
  %``Search for a charged Higgs boson in pp collisions at $ \sqrt{s}=8 $ TeV,''
  JHEP {\bf 1511}, 018 (2015).
  doi:10.1007/JHEP11(2015)018
  [arXiv:1508.07774 [hep-ex]].
  %%CITATION = doi:10.1007/JHEP11(2015)018;%%
  %20 citations counted in INSPIRE as of 17 May 2016

%\cite{CMS:2016szv}
\bibitem{CMS:2016szv} 
  CMS Collaboration [CMS Collaboration],
  %``Search for charged Higgs bosons with the $\mathrm{H}^{\scriptscriptstyle \pm}\rightarrow \tau^{\scriptscriptstyle \pm}\nu_{\tau}$ decay channel in the fully hadronic final state at $\sqrt{s} = 13~\mathrm{TeV}$,''
  CMS-PAS-HIG-16-031.
  %%CITATION = CMS-PAS-HIG-16-031;%%
  %24 citations counted in INSPIRE as of 29 Jun 2018

%\cite{Aaboud:2018gjj}
\bibitem{Aaboud:2018gjj} 
  M.~Aaboud {\it et al.} [ATLAS Collaboration],
  %``Search for charged Higgs bosons decaying via $H^{\pm} \to \tau^{\pm}\nu_{\tau}$ in the $\tau$+jets and $\tau$+lepton final states with 36 fb$^{-1}$ of $pp$ collision data recorded at $\sqrt{s} = 13$ TeV with the ATLAS experiment,''
  arXiv:1807.07915 [hep-ex].
  %%CITATION = ARXIV:1807.07915;%%

%\cite{Aad:2013hla}
\bibitem{Aad:2013hla} 
  G.~Aad {\it et al.} [ATLAS Collaboration],
  %``Search for a light charged Higgs boson in the decay channel $H^+ \to c\bar{s}$ in $t\bar{t}$ events using pp collisions at $\sqrt{s}$ = 7 TeV with the ATLAS detector,''
  Eur.\ Phys.\ J.\ C {\bf 73}, no. 6, 2465 (2013)
%%%%%  doi:10.1140/epjc/s10052-013-2465-z
  [arXiv:1302.3694 [hep-ex]].
  %%CITATION = doi:10.1140/epjc/s10052-013-2465-z;%%
  %117 citations counted in INSPIRE as of 06 Sep 2018



%\cite{Khachatryan:2015uua}
\bibitem{Khachatryan:2015uua} 
  V.~Khachatryan {\it et al.} [CMS Collaboration],
  %``Search for a light charged Higgs boson decaying to $ \mathrm{c}\overline{\mathrm{s}} $ in pp collisions at $ \sqrt{s}=8 $ TeV,''
  JHEP {\bf 1512}, 178 (2015)
%  doi:10.1007/JHEP12(2015)178
  [arXiv:1510.04252 [hep-ex]].
  %%CITATION = doi:10.1007/JHEP12(2015)178;%%
  %48 citations counted in INSPIRE as of 29 Jun 2018

%\cite{CMS:2018ect}
\bibitem{CMS:2018ect} 
  CMS Collaboration [CMS Collaboration],
  %``Search for charged Higgs bosons with the H$^{\pm} \to \tau^{\pm}\nu_\tau$ decay channel in proton-proton collisions at $\sqrt{s}=13~\mathrm{TeV}$,''
  CMS-PAS-HIG-18-014.
  %%CITATION = CMS-PAS-HIG-18-014;%%


%\cite{Abbiendi:2013hk}
\bibitem{Abbiendi:2013hk} 
  G.~Abbiendi {\it et al.} [ALEPH and DELPHI and L3 and OPAL and LEP Collaborations],
  %``Search for Charged Higgs bosons: Combined Results Using LEP Data,''
  Eur.\ Phys.\ J.\ C {\bf 73}, 2463 (2013)
%  doi:10.1140/epjc/s10052-013-2463-1
  [arXiv:1301.6065 [hep-ex]].
  %%CITATION = doi:10.1140/epjc/s10052-013-2463-1;%%
  %142 citations counted in INSPIRE as of 20 Jun 2018


%\cite{Eberhardt:2013uba}
\bibitem{Eberhardt:2013uba} 
  O.~Eberhardt, U.~Nierste and M.~Wiebusch,
  %``Status of the two-Higgs-doublet model of type II,''
  JHEP {\bf 1307}, 118 (2013).
%  doi:10.1007/JHEP07(2013)118
%  [arXiv:1305.1649 [hep-ph]].
  %%CITATION = doi:10.1007/JHEP07(2013)118;%%
  %95 citations counted in INSPIRE as of 26 Oct 2016

%\cite{Ivanov:2010wz}
\bibitem{Ivanov:2010wz} 
  I.~P.~Ivanov,
  %``Properties of the general NHDM. II. Higgs potential and its symmetries,''
  JHEP {\bf 1007}, 020 (2010);
%  doi:10.1007/JHEP07(2010)020
%  [arXiv:1004.1802 [hep-th]].
  %%CITATION = doi:10.1007/JHEP07(2010)020;%%
  %17 citations counted in INSPIRE as of 26 Oct 2016
%\cite{Ivanov:2012ry}
%\bibitem{Ivanov:2012ry} 
  I.~P.~Ivanov and E.~Vdovin,
  %``Discrete symmetries in the three-Higgs-doublet model,''
  Phys.\ Rev.\ D {\bf 86}, 095030 (2012);
%  doi:10.1103/PhysRevD.86.095030
%  [arXiv:1206.7108 [hep-ph]].
  %%CITATION = doi:10.1103/PhysRevD.86.095030;%%
  %14 citations counted in INSPIRE as of 26 Oct 2016
%\cite{Keus:2013hya}
%\bibitem{Keus:2013hya} 
  V.~Keus, S.~F.~King and S.~Moretti,
  %``Three-Higgs-doublet models: symmetries, potentials and Higgs boson masses,''
  JHEP {\bf 1401}, 052 (2014);
%  doi:10.1007/JHEP01(2014)052
%  [arXiv:1310.8253 [hep-ph]].
  %%CITATION = doi:10.1007/JHEP01(2014)052;%%
  %21 citations counted in INSPIRE as of 26 Oct 2016
%\cite{Maniatis:2014oza}
%\bibitem{Maniatis:2014oza} 
  M.~Maniatis and O.~Nachtmann,
  %``Stability and symmetry breaking in the general three-Higgs-doublet model,''
  JHEP {\bf 1502}, 058 (2015);
  Erratum: [JHEP {\bf 1510}, 149 (2015)];
%  doi:10.1007/JHEP10(2015)149, 10.1007/JHEP02(2015)058
%  [arXiv:1408.6833 [hep-ph]].
  %%CITATION = doi:10.1007/JHEP10(2015)149, 10.1007/JHEP02(2015)058;%%
  %9 citations counted in INSPIRE as of 26 Oct 2016
%\cite{Moretti:2015cwa}
%\bibitem{3hdm-uni} 
  S.~Moretti and K.~Yagyu,
  %``Constraints on Parameter Space from Perturbative Unitarity in Models with Three Scalar Doublets,''
  Phys.\ Rev.\ D {\bf 91}, 055022 (2015). 
%  [arXiv:1501.06544 [hep-ph]].
  %%CITATION = doi:10.1103/PhysRevD.91.055022;%%
  %5 citations counted in INSPIRE as of 24 Oct 2016


%\cite{Bento:2017eti}
\bibitem{Bento:2017eti} 
  M.~P.~Bento, H.~E.~Haber, J.~C.~Romão and J.~P.~Silva,
  %``Multi-Higgs doublet models: physical parametrization, sum rules and unitarity bounds,''
  JHEP {\bf 1711}, 095 (2017)
%  doi:10.1007/JHEP11(2017)095
  [arXiv:1708.09408 [hep-ph]];
  %%CITATION = doi:10.1007/JHEP11(2017)095;%%
  %9 citations counted in INSPIRE as of 06 Sep 2018
%\cite{Haber:2018iwr}
%\bibitem{Haber:2018iwr} 
  H.~E.~Haber, O.~M.~Ogreid, P.~Osland and M.~N.~Rebelo,
  %``Symmetries and Mass Degeneracies in the Scalar Sector,''
  arXiv:1808.08629 [hep-ph].
  %%CITATION = ARXIV:1808.08629;%%


\end{thebibliography}
\end{document}